# Imaging work and dissipation in the quantum Hall state in graphene

A. Marguerite[1,†], J. Birkbeck[2,†], A. Aharon-Steinberg[1,†], D. Halbertal[1,#], K. Bagani[1], I. Marcus[1], Y. Myasoedov[1], A. K. Geim[2], D. J. Perello[2,*], and E. Zeldov[1,*]

Topology is a powerful recent concept asserting that quantum states could be globally protected against local perturbations [1,2]. Dissipationless topologically protected states are thus of major fundamental interest as well as of practical importance in metrology and quantum information technology. Although topological protection can be robust theoretically, in realistic devices it is often fragile against various dissipative mechanisms, which are difficult to probe directly because of their microscopic origins. By utilizing scanning nanothermometry [3], we visualize and investigate microscopic mechanisms undermining the apparent topological protection in the quantum Hall state in graphene. Our simultaneous nanoscale thermal and scanning gate microscopy shows that the dissipation is governed by crosstalk between counterpropagating pairs of downstream and upstream channels that appear at graphene boundaries because of edge reconstruction. Instead of local Joule heating, however, the dissipation mechanism comprises two distinct and spatially separated processes. The work generating process that we image directly and which involves elastic tunneling of charge carriers between the quantum channels, determines the transport properties but does not generate local heat. The independently visualized heat and entropy generation process, in contrast, occurs nonlocally upon inelastic resonant scattering off single atomic defects at graphene edges, while not affecting the transport. Our findings offer a crucial insight into the mechanisms concealing the true topological protection and suggest venues for engineering more robust quantum states for device applications.

___________________________

[1]Department of Condensed Matter Physics, Weizmann Institute of Science, Rehovot 7610001, Israel

[2]National Graphene Institute and School of Physics and Astronomy, The University of Manchester, Manchester M13 9PL, UK

[†]These authors contributed equally to this work

[#]Present address: Department of Physics, Columbia University, New York, New York 10027, USA

[*]Corresponding author. Email: david.perello@manchester.ac.uk (D.J.P.); eli.zeldov@weizmann.ac.il (E.Z.)

In recent years, major progress has been made in identifying new topological states of matter [1,2] but the extent to which the topological protection is manifested in realistic systems and the microscopic mechanisms leading to its apparent breakdown remain poorly understood. The quantum Hall (QH) effect is a prime example of a topologically protected state exhibiting quantized dissipationless electron transport. While an extremely high degree of conductance quantization has been achieved in engineered systems in GaAs and in graphene [4], QH devices commonly exhibit small but fundamentally important deviations from the ideal quantized conductance. Various mechanisms undermining the topological protection were explored, including imperfect contacts [5], current-induced breakdown [4], absence of edge equilibration [6], and edge reconstruction [7,8]. Nonetheless, how exactly the dissipation in the QH regime occurs on a microscopic level has eluded direct identification. Here we provide nanoscale imaging of the dissipation processes in the QH state in graphene and reveal the intricate mechanisms compromising the apparent global topological protection.

A superconducting quantum interference device, SQUID-on-tip (SOT) [9] acting as nanothermometer (tSOT) with µK sensitivity [3] and effective diameter of ~50 nm was scanned ~50 nm above high-mobility hBN-encapsulated graphene devices (SI1) at $T = 4.2$ K, while simultaneously employing three modalities (Fig. 1a and SI2): (i) *dc* thermal imaging which maps the local temperature variations $T_{dc}(\boldsymbol{r})$ induced by an externally applied current $I_{dc}$. The current was chopped at ~94 Hz and $T_{dc}(\boldsymbol{r})$ was recorded using a lock-in amplifier. (ii) *ac* thermal imaging: the tSOT is mounted on a quartz tuning fork and vibrates parallel to the sample surface at a frequency of ~35 kHz with $x_{ac}$~8 nm amplitude. The resulting $T_{ac}(\boldsymbol{r}) \cong x_{ac} \partial T_{dc}(\boldsymbol{r})/\partial x$, provides a high sensitivity map of the local temperature gradients [3]. (iii) Scanning gate mode [10] in which a voltage $V_{tg}$ is applied to the tip and the induced variations in the two-probe, $R_{2p}(\boldsymbol{r})$, or four-probe, $R_{xx}(\boldsymbol{r})$, sample resistance is imaged.

Topological protection in an *idealized* integer QH system is manifested by three guiding principles [6]:

(*A*) In the quantum Hall plateau the current flows along ballistic chiral edge channels with no backscattering and no dissipation except at the current contacts.

(*B*) In the plateau transition regions dissipation sets in through the bulk of the system.

(*C*) The topological state and the corresponding quantized conductance is robust against local perturbations and is determined by the bulk Chern number and the bulk-edge correspondence.

Global transport measurements of our devices show common QH characteristics, including conductance quantization (SI3), which are qualitatively consistent with the above principles. Surprisingly, when inspected microscopically, we find these principles to be largely violated. A *dc* current $I_{dc} \cong 1$ µA was injected through a 300 nm wide constriction at the bottom edge of sample *A* (Fig. 1b) and collected at a top-right contact. Figures 1c,d show the resulting $T_{ac}(\boldsymbol{r})$ images at two values of $V_{bg}$ (see SI4 and Movie M1 for range of $V_{bg}$). There are two $V_{bg}$–independent features: a large thermal gradient near the constriction due to heat generated in it and diffused through the substrate, and artificial background signal outlining the sample topography (SI2). Notably, $V_{bg}$–dependent ring-like structures appear along the graphene boundaries, which reveal dissipation through phonon emission from individual atomic defects [11]. When $V_{bg}$ is tuned into the $\nu = -10$ QH plateau (Fig. 1c, $\nu = -10.7$) the dissipation occurs along the bottom edge of the sample in violation of principle *A*. Tuning $V_{bg}$ into the QH plateau transition region ($\nu = -1.46$, Fig. 1d), dissipation is observed primarily along the edges rather than in the bulk (as demonstrated in particular by the absence of thermal rings at the atomic defects along the inner edges of the five square holes) in violation of principle *B*. Moreover, at high filling factors the dissipation occurs predominantly along the downstream chiral flow direction from the constriction (counterclockwise for holes in Fig. 1c) with characteristic decay length of 15 µm, and it is rather independent of whether the $V_{bg}$ is tuned into the QH plateau or not (Movie M1). At lower filling factors (Fig. 1d), the dissipation is greatly enhanced in both downstream and upstream directions with no visible chirality, and extends over the entire length of the edges with no apparent decay (SI4).

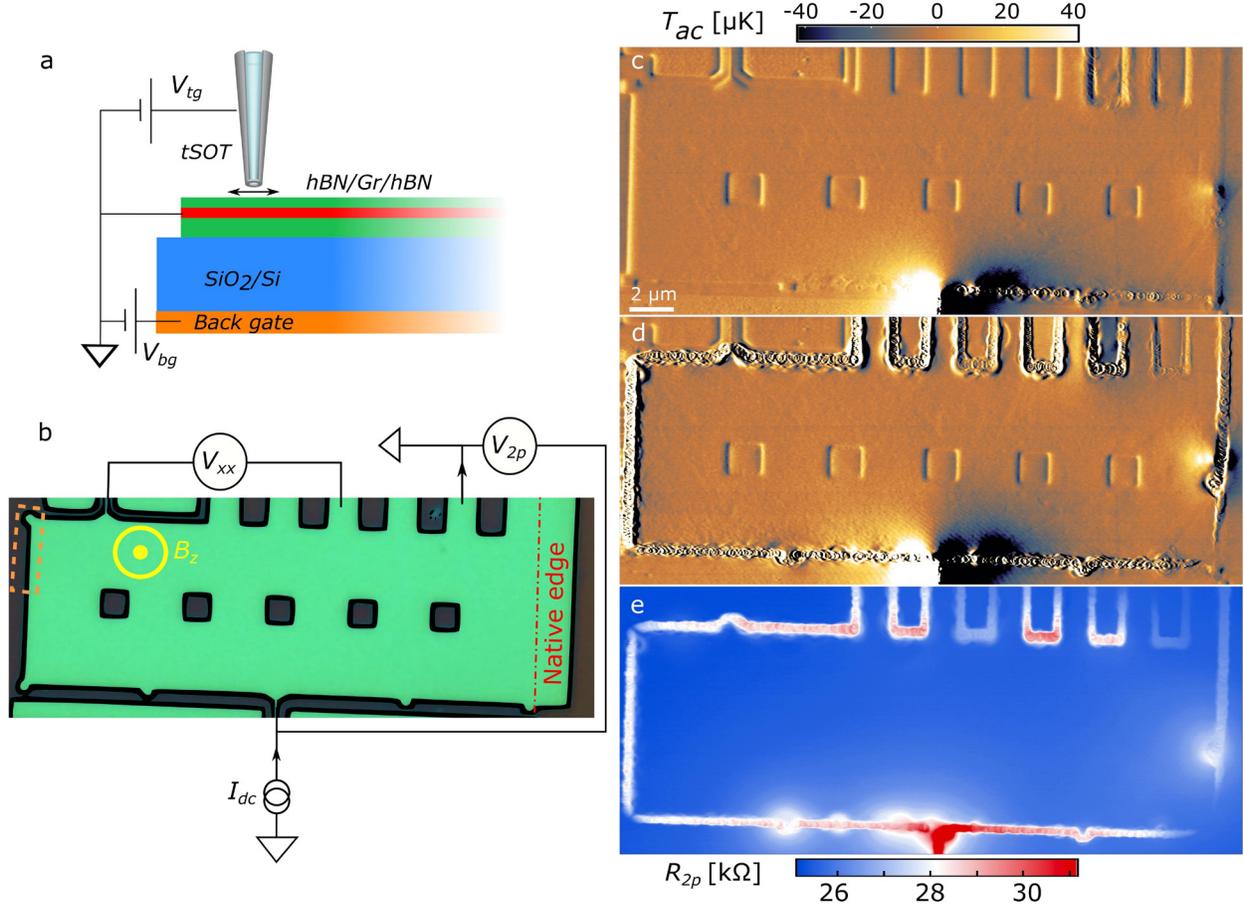

**Figure 1 | Imaging violation of the QH topological protection. a**, Schematic measurement setup with hBN-graphene-hBN heterostructure and scanning tSOT. **b**, Optical image of device *A* patterned with several contacts and five square holes in the center. A *dc* current $I_{dc}$ is driven through the narrow bottom constrictions and drained at the top-right contact (arrows) in presence of applied field $B_z = 1.0$ T at 4.2 K. **c**, $T_{ac}(r)$ thermal image in the vicinity of the $\nu = -10$ QH plateau at $V_{bg} = -6.0$ V ($\nu = -10.7$), $V_{tg} = 8$ V, and $I_{dc} = 1.5$ µA ($R_{2p}I_{dc}^2 = 10$ nW) revealing phonon emission from individual atomic defects in a form of rings along the bottom-right graphene boundary (see Movie M1). **d**, Same as (c) but in the QH plateau transition region at $V_{bg} = -2.5$ V ($\nu = -1.46$) and $I_{dc} = 0.87$ µA ($R_{2p}I_{dc}^2 = 10$ nW), showing enhanced dissipation along all edges with no visible dissipation in the bulk. **e**, Scanning gate $R_{2p}(r)$ image acquired simultaneously with (d) revealing significant tip-induced enhancement of the two-probe sample resistance along the edges.

Lastly, an example of violation of principle *C* is demonstrated in Fig. 1e. Topologically protected states should be robust against local perturbations, and hence positive $V_{tg}$ which depletes holes on a scale much smaller than the sample size should not affect global transport properties. In contradiction, the two-probe resistance $R_{2p}$ of a 30 µm sample is non-trivially affected by a perturbation on a scale of ~50 nm (tip size). The large increase in $R_{2p}(r)$ occurs only along the graphene boundaries and is observed over a wide range of $V_{bg}$ both at QH plateaus (Movie M2) and at plateau transitions (Fig. 1e). Also, importantly, the $R_{2p}(\boldsymbol{r})$ signal is visible along the entire length of the boundaries.

For a closer inspection, we focus on the dashed rectangle in Fig. 1b with a square-shaped protrusion at the top-left corner. The higher resolution $T_{dc}(\boldsymbol{r})$ image (Fig. 2a) reveals disordered heat signal concentrated along two separate contours. The outer contour consists of a series of thermal rings centered along the physical edge of graphene (dashed line). The inner contour with arc-shaped features is visible further inside the sample. Critically, the simultaneously acquired scanning gate $R_{2p}(\boldsymbol{r})$ signal (Fig. 2b) mimics precisely the $T_{dc}(\boldsymbol{r})$ signal along the inner contour while showing no response along the outer contour or elsewhere. This striking difference indicates that the inner and outer contours arise from fundamentally different mechanisms.

To decipher the different mechanisms, consider a diffusive system in steady state in which dissipation is described by local Joule heating, $P(\boldsymbol{r}) = \dot{W}(\boldsymbol{r}) = \boldsymbol{J}(\boldsymbol{r}) \cdot \boldsymbol{E}(\boldsymbol{r}) = \dot{Q}(\boldsymbol{r})$, where power $P$ is the rate of work $\dot{W}$ per unit volume performed by current density $\boldsymbol{J}$ driven by electric field $\boldsymbol{E}$. In this case, the work $W$ is transformed into heat $Q$ locally and hence $\dot{W}(\boldsymbol{r}) = \dot{Q}(\boldsymbol{r})$. Conversely, dissipation in a ballistic system can be highly nonlocal, resulting in $\dot{W}(\boldsymbol{r}) \neq \dot{Q}(\boldsymbol{r})$, as illustrated in Fig. 2c for elastic tunneling through a potential barrier. The work generation $\dot{W}(\boldsymbol{r})$ occurs only where the carriers are accelerated by $\boldsymbol{E}$ within the barrier. Meanwhile heat, $\dot{Q}(\boldsymbol{r})$, or entropy, $\dot{Q}(\boldsymbol{r})/T$, generation processes occur nonlocally as carriers lose their excess kinetic energy remotely via inelastic scattering of phonons far from the initial barrier.

In general, one should consider three main stages – work generation, equilibration through electron-electron scattering, and heat transfer to the environment through phonon emission. The equilibration process due to electron-electron scattering in the QH channels has been extensively studied by spectroscopic transport measurements [12–14]. Such electron-electron scattering results in energy redistribution within the electronic bath, which is undetectable by our technique since no energy transfers to the phonon bath. In the following, we focus on the first stage of $\dot{W}$ generation and the last stage of $\dot{Q}$-release into the phonon bath under steady state conditions, in which the details of the intermediate electron-electron scattering process have no significant effect. In other words, we address the question of where and how the work is generated and where and how the heat is transferred to the environment.

In an ideal current carrying ballistic channel no work generating processes can take place since there is no potential drop along the channel, $E_\parallel(\boldsymbol{r}) = 0$, and hence $\dot{W}(\boldsymbol{r}) = 0$. Paradoxically, however, heat can still be generated by the entropy generating processes, $\dot{Q}(\boldsymbol{r})$. This is the situation at higher $\nu$ where analogous to the tunnel barrier in Fig. 2c, work $\dot{W}(\boldsymbol{r})$ is performed at the bottom constriction in Fig. 1c by injecting energetic charge carriers into the QH edge channels. These chiral carriers flow downstream ballistically and cause nonlocal heating by losing their excess energy to phonons. Since at low temperatures and in absence of disorder, electron-phonon coupling is very weak, phonon emission occurs predominantly through resonant inelastic scattering off single atomic defects along the graphene edges [11]. These defects form quasi-bound states with sharp energy levels that mediate electron-phonon coupling when in resonance with the incoming charge carriers [15,16], giving rise to the $\dot{Q}$ rings observed in Figs. 1c and 2a. Since only forward carrier scattering is allowed in chiral QH channels, phonon scattering does not affect conductivity and is thus invisible in $R_{2p}(\boldsymbol{r})$ image in Fig. 2b and can coexist with full conductance quantization.

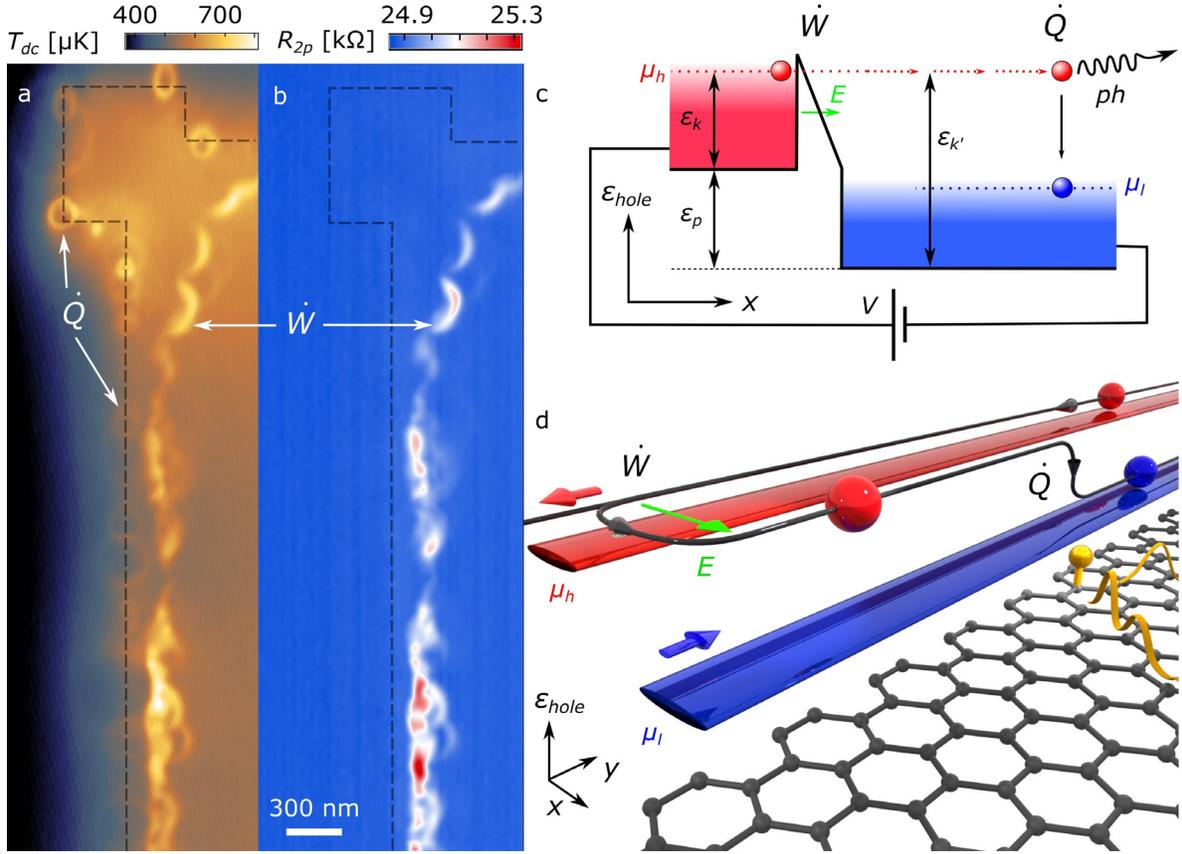

**Figure 2 | Work generating and entropy generating processes. a**, Thermal image $T_{dc}(r)$ in the corner of sample A (dashed rectangle in Fig. 1b) at $V_{bg} = -2.1$ V ($\nu \simeq -1.9$), $V_{tg} = 1.6$ V, and $I_{dc} = 0.62$ μA revealing the $\dot{Q}$ process of phonon emission in form of thermal rings along the graphene boundaries (dashed) and the $\dot{W}$ process of elastic tunneling between QH channels along an inner contour. **b**, Simultaneously acquired $R_{2p}(r)$ image showing $\dot{W}$ process along the inner contour that is identical to (a), while displaying no signal of the $\dot{Q}$ process along the edges. **c**, Schematic diagram of hole-carrier elastic tunneling across a potential barrier in a ballistic regime. The work generation $\dot{W}$ occurs only within the barrier where the electric field $E$ (green) accelerates the carriers transforming the potential energy $\varepsilon_p$ into excess kinetic energy $\varepsilon_{k'}$. The heat $\dot{Q}$ and entropy $\dot{Q}/T$ are generated nonlocally at a remote location through inelastic scattering of phonons. The difference between the high and low electrochemical potentials is provided by the battery, $\mu_h - \mu_l = eV$. **d**, A schematic cartoon of $\dot{W}$ generation by elastic tunneling of hole-carriers (spheres) between counterpropagating QH channels. The transverse electric field $E$ (green) performing the work is caused by the difference in the electrochemical potentials $\mu_h$ (red) and $\mu_l$ (blue). The heat generation $\dot{Q}$ occurs at a remote location through phonon emission (light brown) by inelastic carrier scattering off an atomic defect (yellow) at graphene boundaries.

At lower fillings, however, strikingly different behavior is observed (SI4 and Movie M1). The $\dot{Q}$ rings along the graphene boundaries in Figs. 1d and 2a still reflect nonlocal dissipation, but they are apparently not "powered" by the work generated at the constriction, as evidenced by lack of observable chirality and of signal decay. Instead, the $\dot{W}$ process occurs along the inner contour in Fig. 2b where carriers tunnel elastically between neighboring QH channels with different electrochemical potential $\mu$, as illustrated schematically in Fig. 2d. The tip positioned at $r$ modifies the local separation between the channels by its potential $V_{tg}$. When channels are brought closer, the tunneling rate enhances and the corresponding backscattering current increases by $\delta I_{bs}(r)$, increasing $R_{2p}(r)$ and generating excess local work at a rate of $\delta \dot{W}(r) = \delta I_{bs}(r)(\mu_h(r) - \mu_l(r))/e$. This $\dot{W}(r)$ process, however, is elastic and thus does not generate local $\dot{Q}(r)$ and no phonons are emitted locally (SI11). Instead, the backscattered carriers release their excess energy $(\mu_h(r) - \mu_l(r))$ to phonons elsewhere, predominantly at atomic defects on graphene boundaries. The emitted phonons propagate ballistically increasing the overall sample temperature, including at the instantaneous position of the tip, $r$. Since the resulting overall increase in $\delta T$ and the increase in $R_{2p}(r)$ are both proportional to the tip-induced $\delta \dot{W}(r)$, the $T_{dc}(r)$ signal in Fig. 2a along the inner contour accurately mimics the $R_{2p}(r)$ signal in Fig. 2b, even though no local $\dot{Q}(r)$ is generated (see SI11).

The above picture, however, raises yet another puzzle. In conventional integer QH, backscattering is prohibited by chirality unless in the presence of counterpropagating channels. Such counterpropagating channels are typically only present at the opposite sample edges, whereas the described $\dot{W}(r)$ process requires close proximity between them (Fig. 2d). The attained findings therefore provide microscopic evidence for the presence of edge reconstruction induced by charge (holes) accumulation along the graphene edges as suggested previously [7,8,17–20].

To control this edge reconstruction and investigate its origin, we incorporated plunger gates (PG) as described in Figs. 3a-c, S2, and S3. As detailed in Figs. 3d,e, band bending due to charge accumulation along the edges creates pairs of counterpropagating QH channels which are not topologically protected and exist in addition to the standard topological channels dictated by the bulk-edge correspondence [7,8,19–21]. These "nontopological" channels provide the means for $\dot{W}(r)$ backscattering and work generation along the entire edge of the sample as evidenced in Figs. 1d-e, 2b, and 3f-h. The presence of band-bending-induced nontopological channels is quite insensitive to the bulk $\nu$ and therefore the backscattering $\dot{W}(r)$ occurs for both compressible and incompressible bulk (SI5 and Movie M2). Moreover, since the nontopological pairs are present at both edges of the sample, the $\dot{W}(r)$ and the resulting nonlocal $\dot{Q}(r)$ show no chiral directionality (Figs. 1d-e). The backscattering rate is determined by the separation between the counterpropagating channels which we can tune by PG potential $V_{pg}$. Remarkably, by increasing the hole accumulation the separation between the channels is increased (Figs. 3i,j) leading to elimination of $\dot{W}(r)$ (Fig. 3k) and of the associated nonlocal heating (Figs. 3l,m) in the plunger gate region. Upon further increase of hole accumulation (see SI6, SI7, and Movie M3 for the full sequence), the $\dot{W}(r)$ surprisingly reappears, but at the bulk side of the plunger gate (Figs. 3n-r) where two *copropagating* channels (blue and green in Fig. 3o) are formed and hence no backscattering is naively expected. The green channel, however, creates a closed loop and therefore serves as a backscattering mediator between the downstream $\mu_h$ (red) and upstream $\mu_l$ (blue) channels. Since the green and red channels copropagate along a longer path and are in close proximity due to the steep edge potential (Fig. 3n), the electrochemical potential of the green channel will be close to $\mu_h$. As a result, the overall backscattering rate will be determined by the tunneling rate between the green and blue channels, explaining the dominant $\dot{W}(r)$ signal along this segment. Note that the patterns along this segment (Figs. 3p-r and S7h,i) are smoother, emphasizing the dominant role of edge disorder in the formation of the complex $\dot{W}(r)$ arc-like patterns along the graphene edges (Figs. 1d and 2a,b). Also, since there are almost no atomic defects in the bulk of graphene [11], no $\dot{Q}(r)$ rings are observed along this segment (Movie M3). The $\dot{Q}(r)$ rings are only resolved along the graphene boundaries (Fig. S7), powered by the remote $\dot{W}(r)$, consistent with the observed separation of $\dot{Q}(r)$ and $\dot{W}(r)$ contours in Fig. 2.

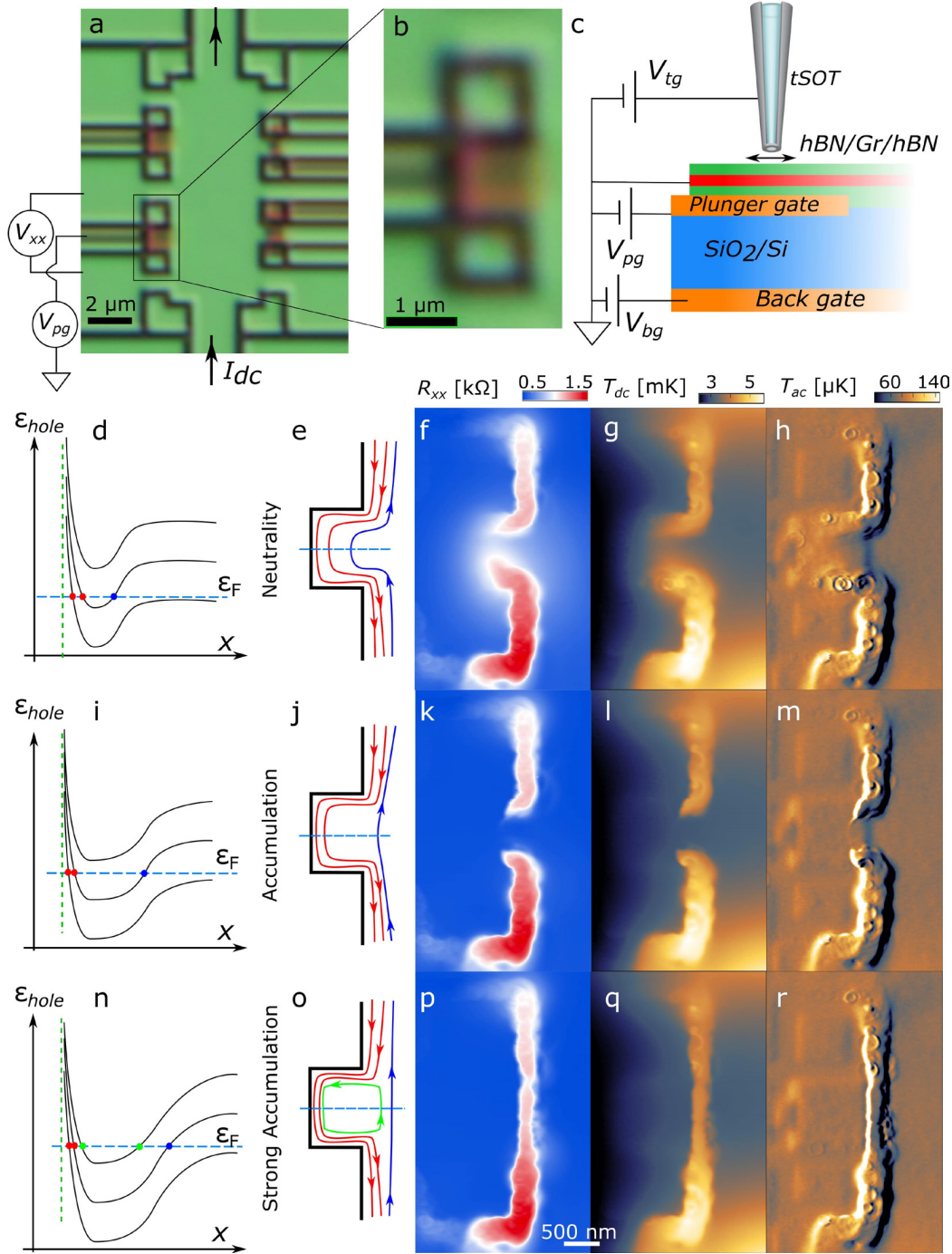

**Figure 3 | Control and separation of the work and entropy generating processes. a**, Optical image of device *B* with four plunger gates and schematic measurement circuit (see Fig. S2). **b**, Zoomed optical image of the plunger gate region showing the etched trenches (dark) and the underlying plunger gate (orange). **c**, Schematic measurement setup of hBN-graphene-hBN heterostructure with a plunger gate and scanning tSOT. **d-h**, Schematic equilibrium LL energy structure for hole charge carriers (**d**), trajectories of QH edge channels (**e**), $R_{xx}(r)$ (**f**), $T_{dc}(r)$ (**g**), and $T_{ac}(r)$ (**h**) in the $\nu = -2$ plateau at $B_z = 0.9$ T, $V_{bg} = -1.2$ V ($\nu = -1.73$), $V_{tg} = 3$ V, $V_{pg} = -0.24$ V (neutral PG), and $I_{dc} = 1.75$ μA. **i-m**, Same as (**d-h**) but for hole accumulating PG ($V_{pg} = -0.4$ V) that suppresses the backscattering $\dot{W}(r)$ process. **n-r**, Same as (**d-h**) but for strongly hole accumulating $V_{pg} = -2.0$ V, which creates $\dot{W}(r)$ contour along the bulk side of the PG.

Even though the plunger gate affects a small region, it significantly influences the global transport (SI3). A positive (hole depleting) $V_{pg}$ cuts off the nontopological pairs, increasing $R_{xx}$ (Fig. S5d) and forcing the current to bypass the PG region through the bulk (SI8). A large tip potential $V_{tg}$ can also cut of the nontopological pairs as described in SI11, SI12, and Movie M5. Note also that the measured $R_{xx}(\boldsymbol{r})$ is essentially current independent (SI9 and Movie M4), ruling out possible current-induced QH breakdown [4].

The edge reconstruction elucidates previously reported discrepancies in the QH state of graphene [8] and other 2DEG systems [21,22]. Although several mechanism were proposed [23,24], edge accumulation was mainly ascribed to electrostatic gating [7,8] which should lead to symmetric accumulation for *p* and *n* doping. We find that at charge neutrality and for both dopings, the accumulation remains hole-type (SI10 and Fig. S10), indicating that the accumulation is predominantly governed by negatively charged impurities. We observe hole-accumulation for both etched and native edges (dashed line in Fig. 1b), despite no chemical exposure for the latter, suggesting that broken bonds at graphene edges become naturally negatively charged. Similar edge accumulation was recently reported in InAs 2DEG [21]. Interestingly, in proximity-induced superconductivity in graphene and InAs 2DEGs, the supercurrent was observed to flow preferentially along the edges [25–28]. Our results may shed light on the underlying mechanism, which is in turn important for studies of topological superconductivity and Majorana physics [29–31]. Note that the upstream edge channels can undermine the apparent topological protection only if the channels are not well equilibrated [6]. We observe the equilibration length of the upstream channels to be in excess of our sample size of 30 μm (SI6), which provides a possible explanation for the difficulty in achieving precise QH quantization in exfoliated graphene devices. Our findings suggest that the detrimental edge reconstruction can be mitigated by passivation or edge-potential engineering [18,32]. The developed concept of simultaneous work and dissipation imaging, combined with their nanoscale control and spectroscopic analysis, provides a novel tool for investigation of microscopic mechanisms of energy loss and scattering in various quantum and topological systems and in operational electronic nanodevices.

**Acknowledgments** The authors thank G. Zhang, I. V. Gornyi, A. D. Mirlin, and Y. Gefen for stimulating discussions and theoretical analysis, M. E. Huber for SOT readout setup, and M. L. Rappaport for technical assistance. This work was supported by the European Research Council (ERC) under the European Union's Horizon 2020 research and innovation program (grant No 785971), by the Israel Science Foundation ISF (grant No 921/18), by the Minerva Foundation with funding from the Federal German Ministry of Education and Research, by the Weizmann–UK Making Connections Program, and by Manchester Graphene-NOWNANO CDT EP/L-1548X. EZ acknowledges the support of the Leona M. and Harry B. Helmsley Charitable Trust grant 2018PG-ISL006.

**Author contributions** A.M., A.A., D.H., I.M., and E.Z. conceived the experiments. J.B. and D.J.P. conceived and fabricated the samples. A.M. and A.A. carried out the measurements and data analysis. D.H. and I.M. performed preliminary studies. K.B. and Y.M. fabricated the SOTs and the tuning fork feedback. A.M., E.Z., A.A., D.J.P., J.B and A.K.G. wrote the manuscript. All authors participated in discussions and writing of the manuscript.

## Supplementary Information

### SI1. Device fabrication

Monolayer graphene heterostructures were fabricated by exfoliating natural graphite and hBN flakes onto oxidized silicon wafers (290 nm of $SiO_2$) and stacking via a polymer stamping method [33]. This method can achieve contamination-free areas limited only by the size of the hBN. Our devices comprised of a relatively thick bottom hBN crystal (>30 nm) with a thinner (10-20 nm) crystal covering the monolayer graphene. We intentionally misaligned the graphene edges with respect to either hBN (>5°) to avoid any superlattice effects. For samples *B* and *C* incorporating a plunger gate, the bottom gate structures were first patterned and metallized with Cr/Au (1 nm/9 nm), followed by transferring of the annealed hBN/Gr/hBN heterostructure.

We used electron beam lithography (Raith EBPG5200) with a bilayer (A3 495K, A3 950K) PMMA mask to define both the contact location and sample geometry. To improve either contact resistance or edge sharpness, we incorporated two different $CHF_3$ and $O_2$ reactive ion etching recipes for each of the contacts and mesa definition. Contacts were defined by mixed chemical/physical etching (5 W RIE, 150 W ICP) to improve hBN selectivity over PMMA, and thus allowing etching and metallization in a single step. Mesa etching incorporated a physical RIE process (20 W RIE, 0 W ICP), followed by a weak $Ar/O_2$ RIE etch to remove the residual exposed graphene step at the edges [27]. Contact metallization was done via e-beam evaporation of Cr/Au (1 nm/70 nm) and standard liftoff procedure. Finally, before scanning, the samples were soaked in a tetramethylammonium hydroxide (TMAH)-based alkaline developer (MIF-319) to remove residual PMMA resist from the surface of the heterostructure. These fabrication procedures are known to produce samples showing high electron mobility and ballistic transport, with a momentum-relaxing mean free path limited by the sample dimensions [11,34].

Sample *A* had a main chamber of 30×10 $\mu m^2$ (Figs. 1b and S1). Constrictions of 300 and 200 nm width at the bottom and top-left edges were designed to allow injection of energetic carriers into the QH edge channels. We also etched a series of 1.5×1.5 $\mu m^2$ holes in the center of the main chamber in order to visualize dissipation in the center of the device by detecting $\dot{Q}$ rings along the inner edges of the holes as demonstrated in [3]. Protruding 500×250 $nm^2$ rectangles on the lower and left edges of the mesa served as poking pads to facilitate tSOT scanning height control using tuning fork feedback. For the native edge (dashed line in Fig. S1), we choose a straight region of the flake, which indicates one of the main crystallographic cleaving directions of graphene. Samples *B* (Figs. 3a, S2 and S5a) and *C* (Fig. S3) were defined by 200 nm wide trenches, with the main chamber size of 10×4 $\mu m^2$ and 13×5 $\mu m^2$ respectively. The typical two-probe resistances in zero magnetic field at 4.2 K in sample *A* were 1 to 2 kΩ and $\simeq$ 8 kΩ through the constriction, and 7 to 8 kΩ in samples *B* and *C*.

The hBN encapsulated graphene samples residing on $Si/SiO_2$ substrates were glued to Au-plated G10 chip carrier using silver paint. Gold wires for electrical contacts were glued to lithographically defined bonding pads using silver epoxy cured at room temperature for 4 hours. The scanning probe microscope with the sample reside in a brass vacuum chamber that is immersed in liquid He at 4.2 K. The chamber is filled with He exchange gas at $\simeq$ 60 mbar pressure providing thermal coupling between the tSOT and the sample [3] and a good thermal contact between the entire microscope and the liquid He bath.

In the $\dot{Q}$ process phonons are emitted from atomic defects and propagate ballistically throughout the sample. The measured local temperature increase reflects the local energy density of the excess phonons. In the microscopic vicinity of the phonon emitter the excess temperature is determined by the emission rate and the distance from the emitter [3,11], and is essentially independent of the thermal resistance between the sample and the environment. The latter determines the overall average temperature increase of the entire sample dictated by the overall power dissipated in the sample, which will appear as a small global increase in $T_{dc}$.

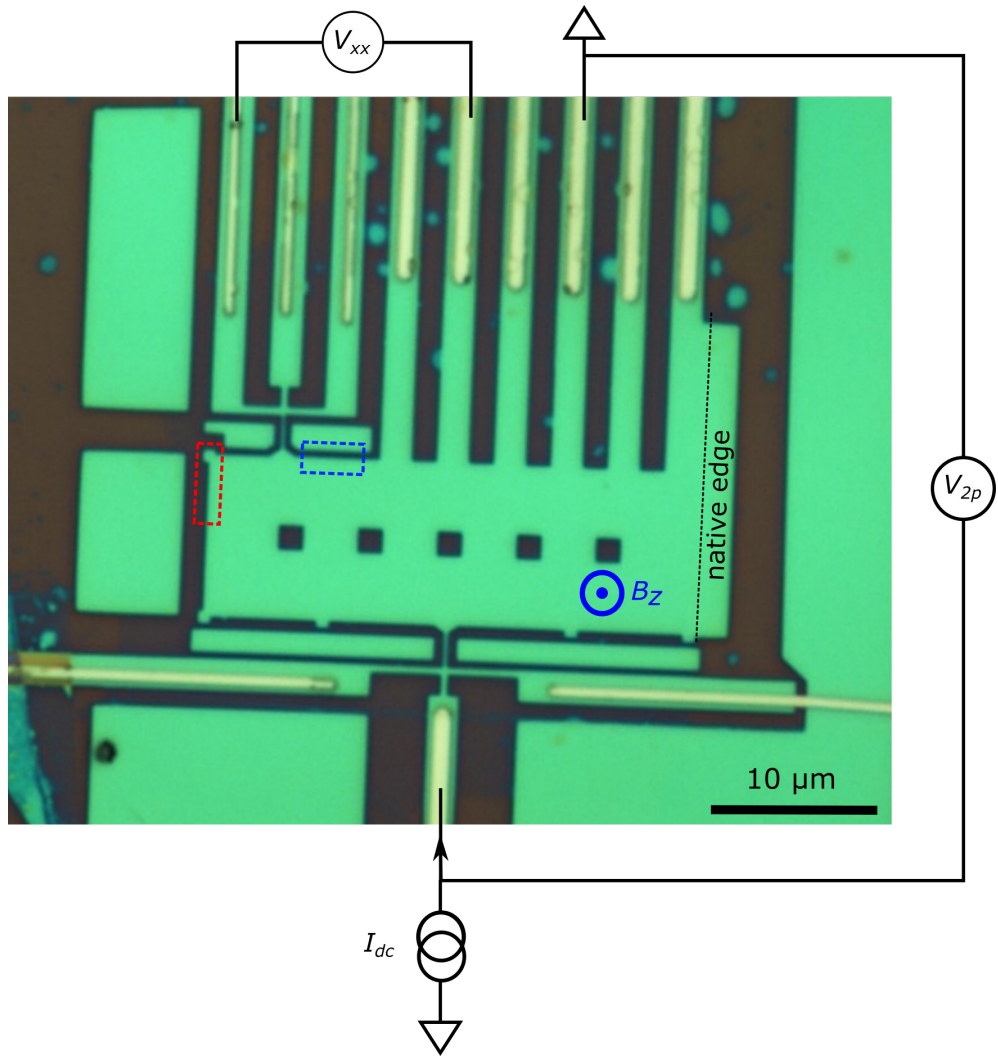

**Figure S1 | Optical image of sample A.** Shown are hBN/graphene/hBN heterostructure (green), etched regions exposing the SiO$_2$/Si substrate (dark), and the metal contacts (yellow). The dashed rectangles mark the regions presented in Fig. 2 (red) and movie M2 (blue). The current is applied to the bottom constriction and drained at the top contact, and the corresponding voltages $V_{xx}$ and $V_{2p}$ are measured in the scanning gate mode. The dashed line on the right shows the native edge of graphene encapsulated in hBN.

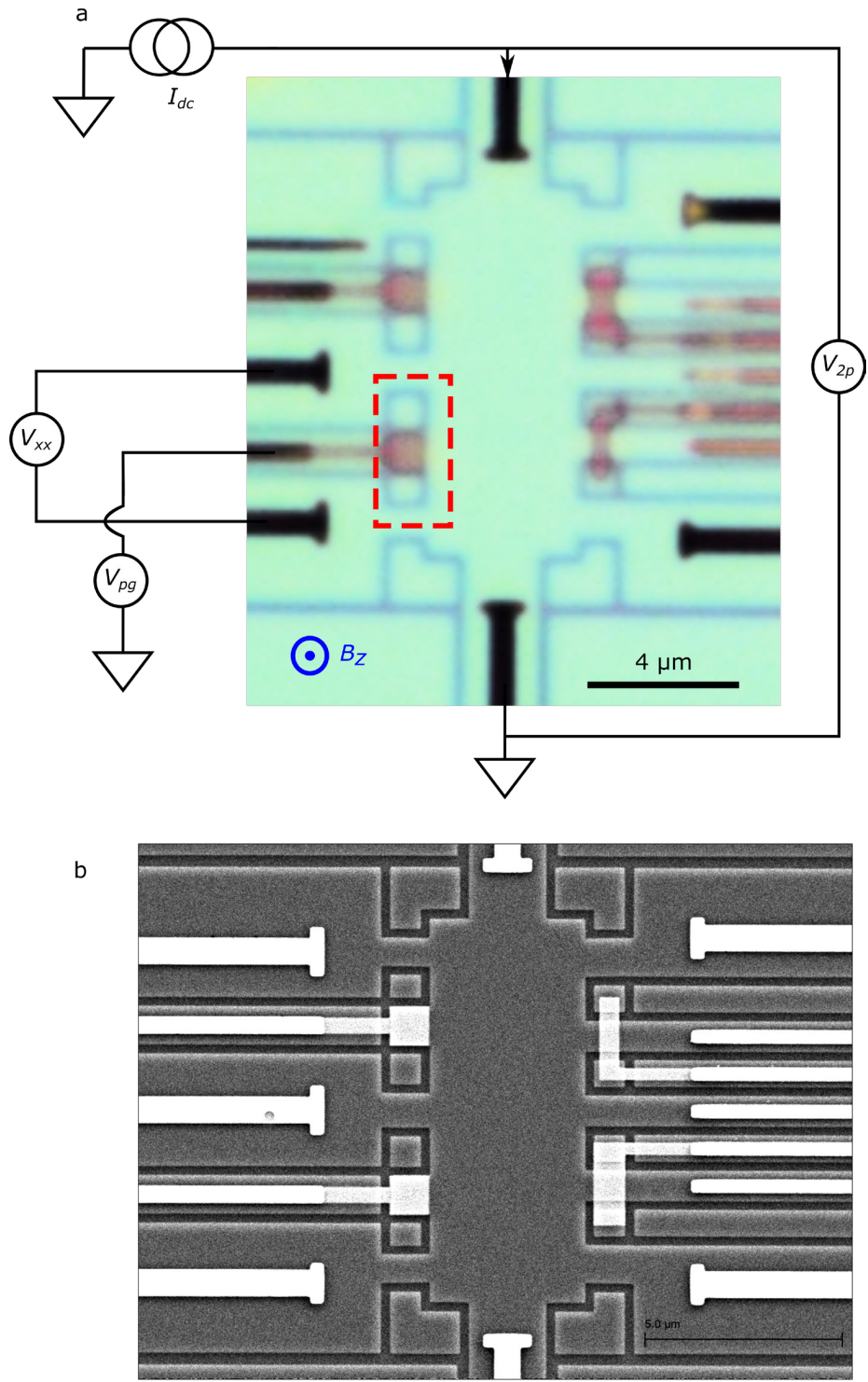

**Figure S2 | Optical and SEM images of sample B. a**, Optical image showing hBN/graphene/hBN heterostructure (light green), etched tranches exposing the SiO₂/Si substrate (light blue), bottom plunger gates (light brown), and the metal contacts (dark). The dashed rectangle marks the region presented in Figs. 3, S7, S9, S10, S12 with variable voltage $V_{pg}$ applied to the PG. The current is applied to the top contact and drained at the bottom contact and the corresponding voltages $V_{xx}$ and $V_{2p}$ are measured in the scanning gate mode. **b**, SEM micrograph of a twin sample of device B showing (from bright to dark) the metal contacts, four plunger gates, hBN/Gr/hBN, and the etched trenches.

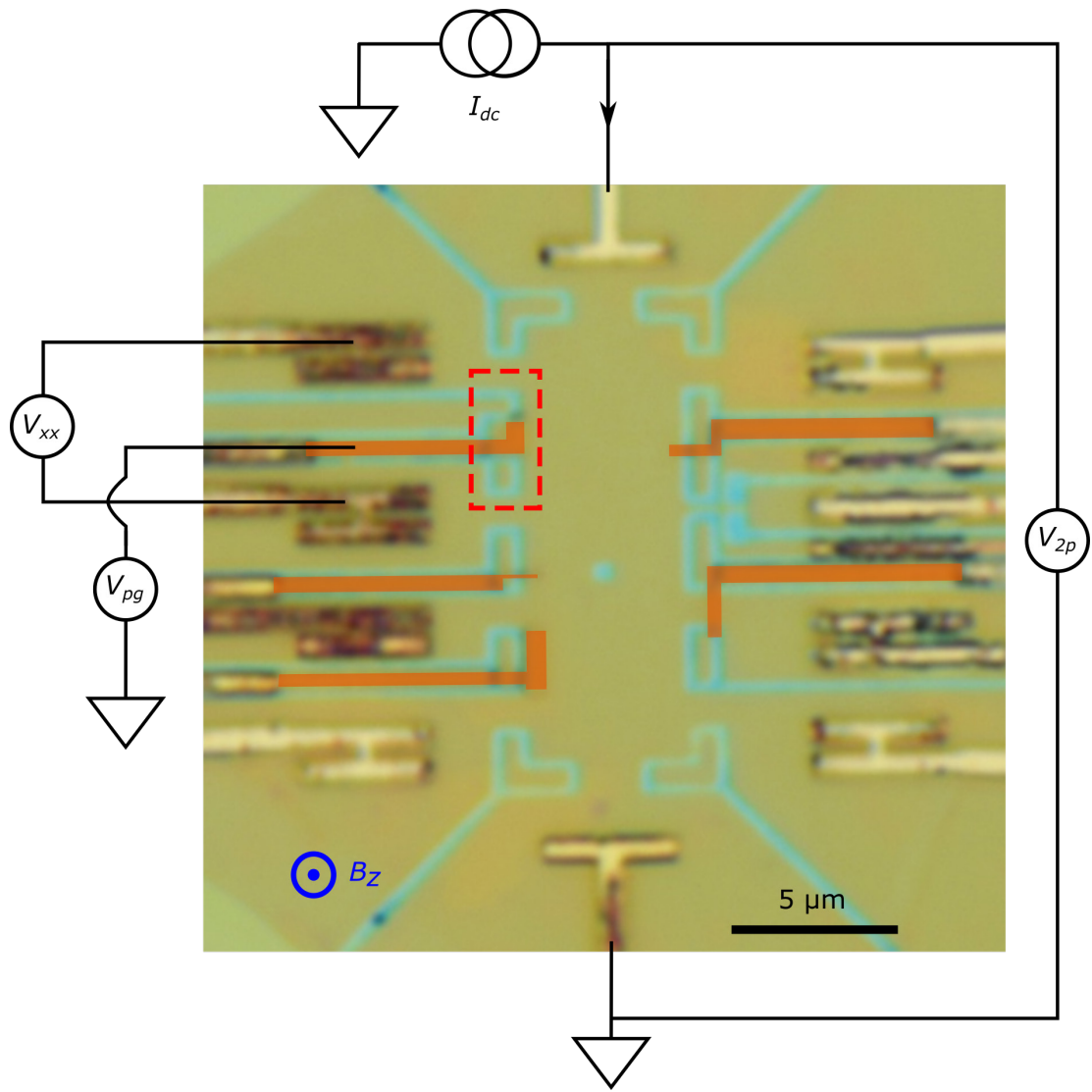

**Figure S3 | Optical image of sample C**. Shown are hBN/graphene/hBN heterostructure (light brown), etched trenches exposing the SiO$_2$/Si substrate (light cyan), and the metal contacts (yellow). The bottom plunger gates are hard to distinguish in the optical image and are artificially highlighted with dark orange color. The dashed rectangle marks the region presented in Fig. S8. The current is applied to the top contact and drained at the bottom and the corresponding voltages $V_{xx}$ and $V_{2p}$ are measured in the scanning gate mode.

## SI2. tSOT fabrication and measurement schemes

The Pb tSOTs were fabricated as described in Ref. [9] with effective diameters of 43, 57 and 50 nm for samples *A*, *B*, and *C* respectively. All the presented scanning studies were carried out at 4.2 K in presence of He exchange gas at ~60 mbar pressure in magnetic fields of $B_z = 1$ T for sample *A*, 0.9 T for sample *B* and 1.2 T in sample *C*. At these conditions the tSOTs displayed thermal noise of down to 0.75 µK/Hz$^{1/2}$. For height control the tSOT was attached to a quartz tuning fork as described in Refs. [35,36] and was electrically excited at the resonance frequency of ~35 kHz. The scanning was performed at a constant height of 25 to 50 nm above the surface of the top hBN. The tuning fork was vibrated along the $\hat{x}$ direction, causing the tSOT to vibrate with it with a controllable amplitude $x_{ac}$ ranging between 5 to 30 nm rms. The typical scanning parameters were: pixel size 6 to 30 nm, acquisition time 10 to 60 ms/pixel, with image sizes from 100 × 100 to 500 × 500 pixels/image. The imaging was performed using four modalities:

*dc thermal imaging $T_{dc}(\mathbf{r})$*

A current $I_{dc}$ is applied to the sample chopped by a square wave at a frequency of 94 Hz and the corresponding thermal map $T_{dc}(\mathbf{r})$ of the sample is acquired using lock-in amplifier locked to the chopping frequency. As a result, the $T_{dc}(\mathbf{r})$ image provides a map of the current-induced local temperature increase in the sample.

*Second harmonic thermal imaging $T_{2f}(\mathbf{r})$*

Similar to the $T_{dc}(\mathbf{r})$, instead of the *dc* current, a sinusoidal *ac* current $I_{ac}$ is applied to the sample at frequency $f$ and the corresponding thermal map $T_{2f}(\mathbf{r})$ of the sample is acquired by lock-in amplifier locked to the second harmonic frequency $2f$.

*ac thermal imaging $T_{ac}(\mathbf{r})$ at tuning fork frequency*

Utilizing the fact that the tSOT is mounted on the tuning fork, we also measure $T_{ac}(\mathbf{r}) \cong x_{ac} \partial T_{dc}(\mathbf{r})/\partial x$ by using a lock-in amplified locked onto the excitation frequency of the tuning fork. The advantage of this mode is that it enhances the visibility of the sharp local features relative to the smooth background, and in particular of the $\dot{Q}(\mathbf{r})$ rings. The $T_{ac}(\mathbf{r})$ signal, however, contains another small component due to the few mK self- heating of the tSOT induced by the measurement current applied to it. In absence of a current in the sample, the self-temperature of the tSOT is slightly position dependent due to the cooling of the tip by the sample (see Fig. S5e in Ref. [3] and its accompanying description) which results in a small contrast between the hBN regions in the sample and the etched trenches. The $T_{ac}(\mathbf{r})$ contains a contribution from the gradient of this contrast, visible as contour lines of the etched hBN mesas in Figs. 1c,d. These contours, which are visible also in the absence of a current in the sample, are highly beneficial for navigation and visualization of the sample with high precision. Since this signal is current independent, it is absent in the $T_{dc}(\mathbf{r})$ images which measure only the temperature increase due to the chopped current $I_{dc}$.

*Scanning gate imaging $R(\mathbf{r})$*

By applying a voltage $V_{tg}$ between the tSOT and the sample we carry out scanning gate imaging similar to the reported methods [10,37–44] simultaneously with the thermal imaging. In particular, the voltage difference $V$ between a pair of sample contacts is measured using a lock-in amplifier locked to the chopping frequency of the current $I_{dc}$, and then $R(\mathbf{r}) = V(\mathbf{r})/I_{dc}$ is plotted vs. the tip location $\mathbf{r}$. In this manner either two-probe, $R_{2p}(\mathbf{r})$, or four-probe, $R_{xx}(\mathbf{r})$, are attained.

## SI3. Transport characteristics

Four-point transport characterization measurements were performed using standard lock-in techniques at 5.4 Hz. Figure S4 shows the Landau fans of samples B and C. From the slopes of the resistance minima we extract the capacitance $C = \frac{e\nu}{\phi_0}\frac{\partial B_z}{\partial V_{bg}}\Big|_{\nu\in\{\pm 2,\pm 6,\pm 10,...\}} = 1.005\cdot 10^{-8}$ F/cm² $= 6.27\cdot 10^{10}$ e/cm²V for sample B and $0.85\cdot 10^{-8}$ F/cm² for sample C. Using the zero field resistivity data $\rho_{xx}$ we derive the mobility $\mu = 1/(n_e e \rho_{xx}) = 7.0\cdot 10^5$ cm²/Vs and the mean free path $l_{mfp} = \frac{1}{2k_F\rho_{xx}}\frac{h}{e^2} = 4.5$ μm at $n_e = 2.8\cdot 10^{11}$ cm⁻² for sample B, and $\mu = 2.13\cdot 10^5$ cm²/Vs and $l_{mfp} = 1.19$ μm at $n_e = 2.4\cdot 10^{11}$ cm⁻² for sample C. Here $n_e = C(V_{bg} - V_{bg}^{CNP})/e$ is the carrier density, $\nu = n_e\phi_0/B_z$ is the filling factor, $V_{bg}^{CNP} = -0.6$ V is the charge neutrality point (CNP) ($-1.85$ V for sample C), $k_F = \sqrt{\pi n_e}$ is the graphene Fermi wave-vector, and $\phi_0 = h/e$. Because of its unconventional geometry and limited working contacts, we could not measure properly the Landau fan diagram of sample A to extract its mobility and mean free path. However, $R_{xx}$ measurements show a similar behavior from which we extract $V_{bg}^{CNP}$ and the approximate filling factors. Note that Figs. 1 and 2 were acquired at different cool downs resulting in a shift in $V_{bg}^{CNP}$.

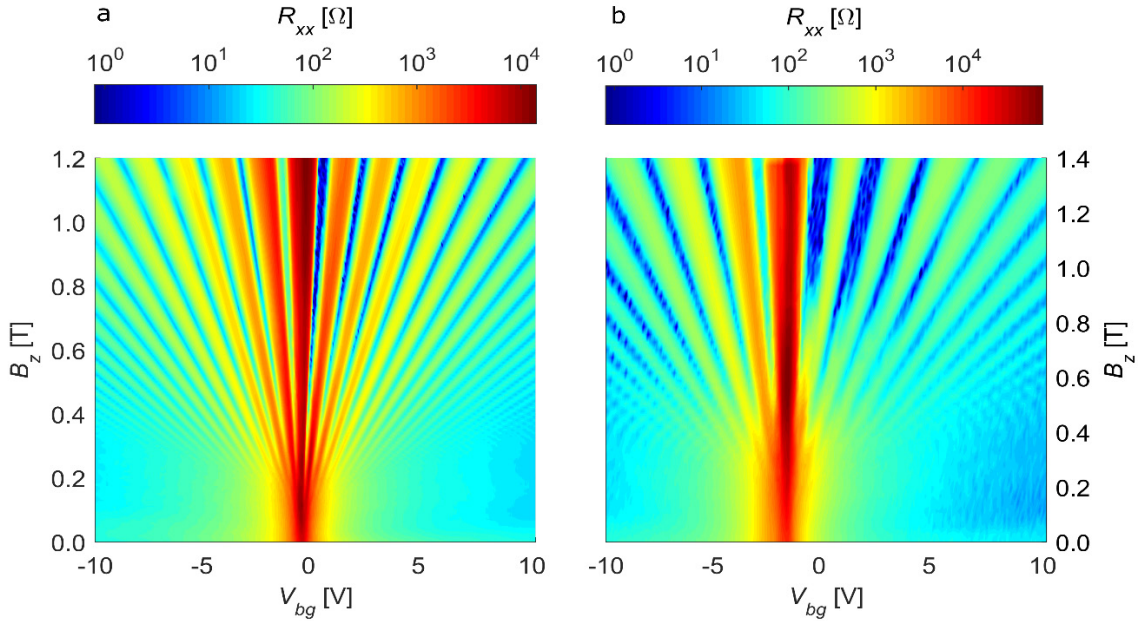

**Figure S4 | Transport measurements of samples B and C.** Color rendering of $R_{xx}$ of samples B (**a**) and C (**b**) as a function of the backgate voltage $V_{bg}$ and the applied perpendicular magnetic field $B_z$.

Figure S5 describes the effect of plunger gate on the global transport. Since the size of the plunger gate is much smaller than the sample, it should naively have no measureable effect in a topologically protected state. Figure S5b shows that when analyzed on a linear scale the variations in $\sigma_{yx}$ and $R_{xx}$ with $V_{pg}$ may not appear to be very significant, however, on a logarithmic scale (Fig. S5c) variations of up to two orders of magnitude in $R_{xx}$ are visible, in particular around $\nu = \pm 2$ plateaus. Figure S5d shows $R_{xx}$ at $V_{bg} = -1$ V in the vicinity of $\nu = -2$ plateau vs. $V_{pg}$ displaying a stepwise increase from $R_{xx} \simeq 100$ Ω to over 4 kΩ for $V_{pg} \geq -0.23$ V. Hole edge accumulation creates additional pairs of counterpropagating nontopological channels that reduce the sample resistivity. When the hole accumulation is depleted by applying a positive $V_{pg}$ the highly conductive nontopological channels are cut off (Fig. S9a) leading to the observed sharp increase in $R_{xx}$. Hole edge accumulation is present also for *n* doping of graphene and is visible in Fig. S5c up to $V_{bg} \cong 2.3$ V above which the $V_{pg}$ dependence decreases significantly, indicating the dominant contribution of negatively charged impurities to the hole edge accumulation (see SI10).

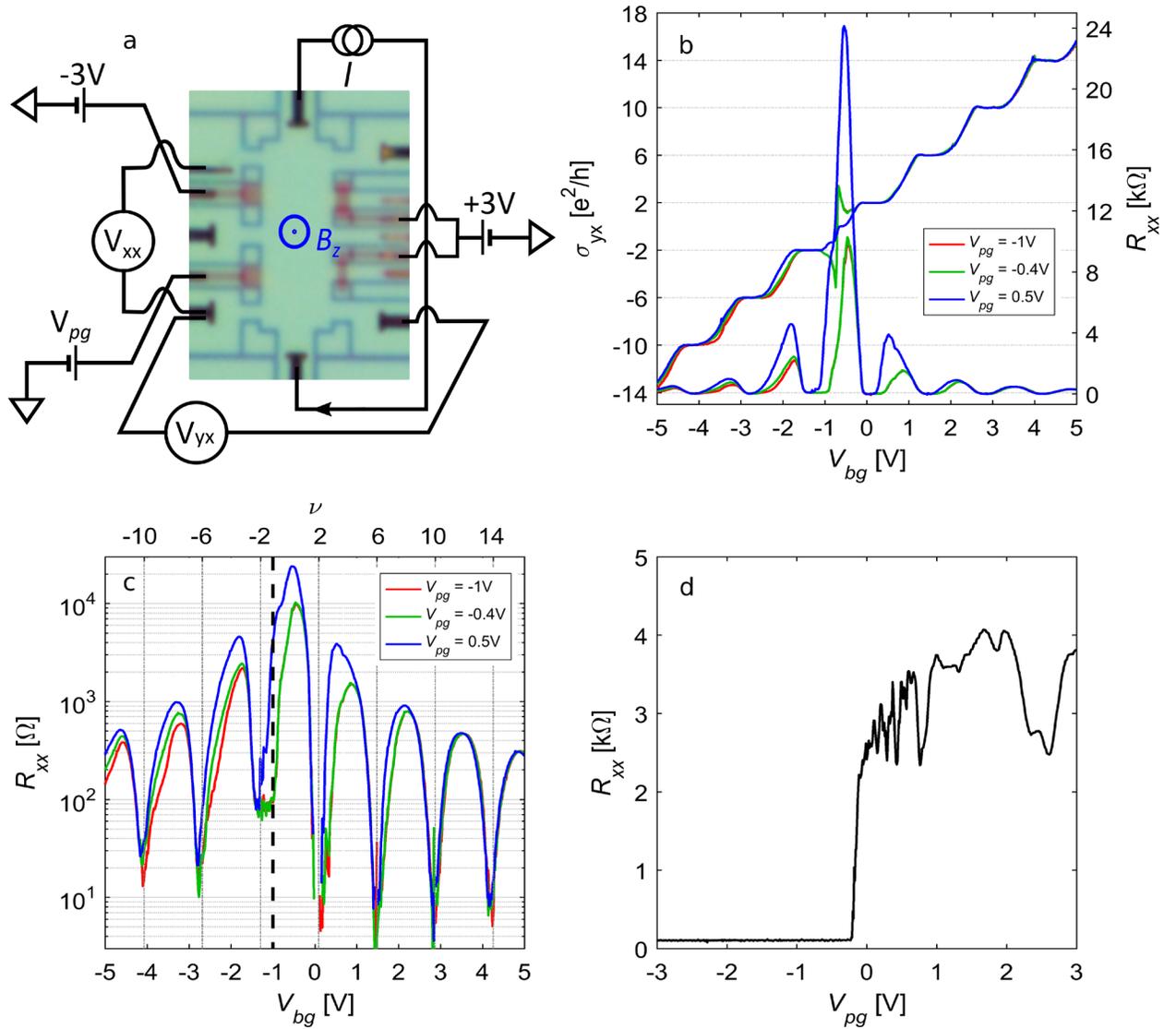

**Figure S5 | Effect of the plunger gate on transport characteristics. a**, An optical image of device *B* with the measurement scheme. **b**, Four probe measurements of $\sigma_{yx}$ and $R_{xx}$ vs. back gate voltage $V_{bg}$ for different $V_{pg}$ values at $B_z = 0.9$ T and $I_{ac} = 50$ nA at 93.72 Hz. A voltage of 3 V was applied to the two plunger gates on the right edge and $-3$ V to the top plunger gate on the left edge. In this configuration the nontopological QH channels on the right edge of the sample are cut off giving rise to enhanced $R_{xx}$ response on the left edge upon varying $V_{pg}$. **c,** $R_{xx}$ vs. $V_{bg}$ from (b) plotted on a logarithmic scale. **d,** Four probe measurement of $R_{xx}$ vs. $V_{pg}$ at $V_{bg} = -1$ V (dashed line in (c)).

## SI4. Movie M1 – Thermal imaging $T_{ac}(r)$ vs. $V_{bg}$ of sample *A*

The full sequence of $T_{ac}(r)$ scans shown in Figs. 1c,d is presented in movie M1 for backgate voltages $V_{bg}$ incremented from $-8$ V to 8 V at 4.2 K, $B_z = 1$ T, and $V_{tg} = 8$ V. A current $I_{dc}$ is driven from the bottom constriction to one of the top contacts (see Figs. 1b and S1) and the value of the current is adjusted with $V_{bg}$ to maintain total power dissipated in the sample of $R_{2p}I_{dc}^2 = 10$ nW (current is ranging from 0.4 µA to 2.1 µA). The chirality of the system is counterclockwise for negative LLs and clockwise for positive LLs. In the movie, one can observe the evolution of entropy generation processes $\dot{Q}$, visible as sharp rings along the edges, and the evolution of work generation processes $\dot{W}$, which appear in the form of larger more blurred features. At large filling factors $|\nu| \geq 10$, predominantly downstream $\dot{Q}$ rings are visible along the bottom edge of the sample to the right (left) of the constriction for negative (positive) $\nu$. In this case the number of downstream channels is significantly larger than of the upstream edge-reconstructed channels. As a result, the channels are better equilibrated and hence there is less backscattering and less work performed along the edges. In this situation most of the work is performed at the constriction and the energetic carriers injected at the constriction flow downstream and lose their excess energy through resonant phonon emission at the atomic defects visible as the $\dot{Q}$ rings [11]. These rings decay over a distance of ~15 µm from the constriction. At $|\nu| \lesssim 10$, $\dot{W}$ arcs begin to appear in addition to the $\dot{Q}$ rings along both downstream and upstream directions and the chirality is gradually lost. This behavior originates from backscattering between counterpropagating nontopological channels resulting in work generation along the channels giving rise to $\dot{W}$ arcs. This work, generated along the entire length of the channels rather than at the constriction, in now the dominant energy source that "feeds" the $\dot{Q}$ rings, explaining the absence of decay in the ring intensity and the absence of chirality (see discussion in SI6). This dissipation, distributed over the full length of the edges, becomes most prominent in the lowest LL, $n_{LL} = 0$, where no topological edge channels are present. Yet most of the current still flows along the edges due to the presence of one or more pairs of counterpropagating nontopological edge channels. In this metallic state, as well as in higher LL metallic states, instead of the commonly assumed backscattering between the opposite edges of the sample, most of the backscattering occurs between the counterpropagating channels within the edges. This is the reason that in Movie M1, we hardly observe any dissipation in the bulk at any value of $V_{bg}$, except very close to CNP, where the overall dissipation in the sample reaches a maximum revealing barely visible $\dot{Q}$ rings along the inner edges of the square holes ($\nu = -0.14$ frame in M1).

The right boundary of the sample is a native edge along one of the main crystallographic cleaving directions of graphene encapsulated in hBN (dashed line in Fig. S1). We note that in general the $T_{ac}(\boldsymbol{r})$ signal along this edge does not differ qualitatively from the signal along the other reactively ion milled edges (section SI1). In the lower-right corner of the sample two diagonal line defects are visible (see two faint diagonal lines in $\nu = 18.34$ frame and array of faint rings along the lower diagonal defect at $\nu = 1.18$). The dissipation along the native edge changes its character upon crossing these line defects, showing some hot spot formation at the intersections.

### SI5. Movie M2 – Scanning gate imaging $R_{xx}(\boldsymbol{r})$ vs. $V_{bg}$ in sample A

A sequence of scanning gate images of the four-probe resistance $R_{xx}(\boldsymbol{r})$ in a zoomed-in region along the top boundary of sample A (blue dashed rectangle in Fig. S1) is presented in Movie M2. The $R_{xx}(\boldsymbol{r}) = V_{xx}(\boldsymbol{r})/I_{dc}$ is recorded as a function of the tip position $\boldsymbol{r}$ for a range of back gate voltages $V_{bg}$ from $-6$ V to $-1.9$ V ($\nu$ from $-10.7$ to $0.09$) in small steps of ~30 mV. Here $B_z = 1.0$ T, $V_{tg} = 1$ V, and the value of the applied current $I_{dc}$ is adjusted for each $V_{bg}$ to maintain total power dissipated in the sample of $R_{2p}I_{dc}^2 = 2$ nW. The dashed horizontal line denotes the top edge of the sample.

The movie reveals a number of important observations. The small positive $V_{tg}$ induces a small local perturbation by slightly reducing the local density of holes, decreasing the distance between the edge channels, thus increasing the backscattering and $\dot{W}$. The tip-induced increase in $R_{xx}(r)$ occurs predominantly along the edge of the sample revealing the location of $\dot{W}$ scattering between the channels. The patterns are highly disordered and often appear to be discontinuous indicating large disorder-induced variations in the position and the number of nontopological pairs of channels. The $R_{xx}(r)$ increase is observed over the entire range of the $V_{bg}$ values spanning from $\nu = -10.7$ to $0.09$. This shows that the edge reconstruction and the backscattering are present for both incompressible and metallic bulk states preventing exact quantization and full apparent topological protection. Near CNP the tip-induced $R_{xx}(r)$ enhancement can be very large reaching values in excess of 10 kΩ.

It is important to note that the additional nontopological channels do not break the true topological state but rather conceal it. This may sound paradoxical since topological state must be robust against local perturbations. The nontopological channels, however, extend over the entire length of the edges, hence do not constitute "local perturbations". By providing additional parallel conductance paths, the nontopological channels compromise the "apparent" topological protection by rendering conductance values that deviate from the exact quantization and give rise to work and dissipation along the entire edges. Since the additional channels are nontopological, their conductance is prone to local disorder and to the tip- and plunger-gate induced changes.

A close inspection of the movie uncovers another interesting feature. The average distance of the contour of enhanced $\delta\dot{W}$ from the edge of the sample varies periodically with $V_{bg}$. Although hard to resolve, in the first three frames of the movie the contour shifts away from the edge, followed by gradual approach towards the edge upon increasing $V_{bg}$ from $-5.9$ to $-4.52$ V. At $V_{bg} = -4.48$ V, however, an additional contour starts to develop further away from the edge, becoming clearly resolved at $-4.42$ V, followed by fractured coexistence of the two contours. From $V_{bg} = -4.35$ V a single contour gradually intensifies and approaches the edge until at $-3.00$ V a new contour away from the edge reemerges, followed again by disappearance of the original contour and gradual movement towards the edge with $V_{bg}$. The emergence of the additional contours occurs over a short span of $V_{bg}$ in the vicinity of the $\nu = -10, -6,$ and $-2$ plateaus, but does not coincide with the plateaus similar to the conclusions of Ref. [8].

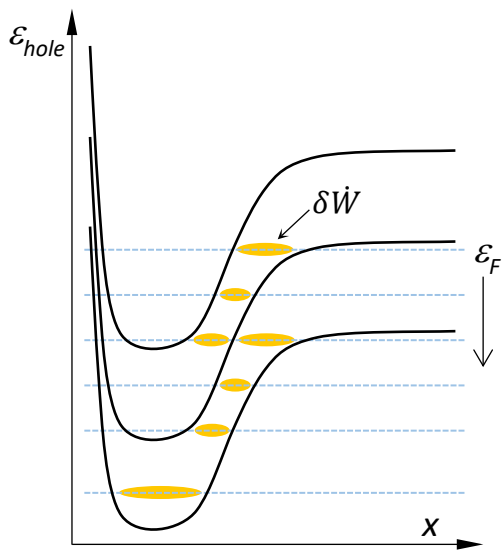

**Figure S6 | Evolution of work generation contours with back gate $V_{bg}$.** Simplified schematic equilibrium diagram of the hole LLs bending (black) due to edge accumulation. With lowering $\varepsilon_F$ (dashed blue lines), the contour of enhanced backscattering $\delta\dot{W}$ (yellow oval) moves towards the edge. Upon crossing the next bulk LL, a new $\delta\dot{W}$ contour appears and moves toward the edge with decreasing $\varepsilon_F$.

This periodic appearance of the contours followed by gradual movement towards the edge can be qualitatively understood by referring to the schematics in Fig. S6, showing the hole LLs bent due to the edge accumulation along with several positions of Fermi level $\varepsilon_F$ controlled by $V_{bg}$. For the highest $\varepsilon_F$ three downstream and two upstream QH channels are present. This situation is qualitatively analogous to the strong accumulation case induced by the plunger gate in Figs. 3n-r. Due to the steep edge potential, the three closely squeezed downstream channels are expected to be well equilibrated whereas the two more remote upstream channels should have larger difference between their electrochemical potentials. As a result, the tip-induced change in the scattering rate $\delta \dot{W}$ should be most pronounced at the location of the inner channels (yellow oval) as indeed observed in the plunger gate example in Figs. 3p-r. Upon decreasing $\varepsilon_F$ the location of $\delta \dot{W}$ contour will shift towards the edge until $\varepsilon_F$ reaches the next bulk LL. In this case an additional inner $\delta \dot{W}$ contour should appear (two ovals in Fig. S6) giving rise to the observed periodic emergence and movement of the contours towards the edge with decreasing $\varepsilon_F$.

## SI6. Movie M3 – Plunger gate $V_{pg}$ dependence

Movie M3 shows an example of the evolution of the simultaneously acquired thermal $T_{ac}(\boldsymbol{r})$ and scanning gate $R_{xx}(\boldsymbol{r})$ images in sample B upon varying $V_{pg}$ similarly to Fig. 3 for $V_{bg} = -1.05$ V, $V_{tg} = 6$ V, and $I_{dc} = 1.75$ µA. At this higher $V_{tg}$ the $\dot{Q}$ rings and the $\dot{W}$ arc-like features are readily resolved. The $\dot{Q}$ rings due to phonon emission at the atomic defects are observed in the $T_{ac}(\boldsymbol{r})$ images along the entire graphene perimeter, visible in the form of smaller diameter sharp rings. They are powered by the remote $\dot{W}$ process even when the latter are shifted significantly away from the edges by the plunger gate potential. The $\dot{Q}$ rings are invisible in the $R_{xx}(\boldsymbol{r})$ images since the $\dot{Q}$ dissipation processes do not cause carrier back scattering. The larger $\dot{W}$ arc-like features are clearly visualized in the $R_{xx}(\boldsymbol{r})$ images (light blue to red) revealing the work generation through carrier backscattering. Since the work $\dot{W}$ causes nonlocal heating, these features are also observed in $T_{ac}(\boldsymbol{r})$ in a form of halos along their outer contours (note that $T_{ac}(\boldsymbol{r})$ reflects temperature gradient along $x$ direction).

The movie starts with positive $V_{pg}$ which cuts off the nontopological pairs of edge channels (see Fig. S9a schematics and transport data in Fig. S5d) forcing the current to flow through the bulk in the cut off region (diffused light red blob for $V_{pg} = 0.6$ V down to $\sim -0.23$ V). At $V_{pg} \cong -0.23$ V the plunger gate is neutral (Fig. S5d) like depicted schematically in Fig. 3e. Upon further decrease of $V_{pg}$, the induced hole accumulation results in enlarged spatial separation between the counterpropagating channels (Fig. 3j) causing a large suppression of $\dot{W}(\boldsymbol{r})$ in the plunger gate region. For $V_{pg} \lesssim -0.6$ V, however, the $\dot{W}(\boldsymbol{r})$ signal reappears, but now on the bulk side of the plunger gate due to appearance of an additional closed loop edge channel (green in Fig. 3o). This loop channel acts as a mediator between the downstream (red) and upstream (blue) channels providing the means for carrier backscattering.

Note that the diameter of the $\dot{Q}$ rings depends both of the tip potential $V_{tg}$ (Movie M5) and on the back gate potential as described in [11]. In Movie M3 $V_{tg}$ and $V_{bg}$ are kept constant and therefore the $\dot{Q}$ rings remain fixed along the graphene boundaries outside the PG. In the PG region, in contrast, the diameter of the rings changes upon varying the $V_{pg}$ as expected [11].

Movie M3 clearly demonstrates the separation of $\dot{W}(\boldsymbol{r})$ and $\dot{Q}(\boldsymbol{r})$ processes and that while the inelastic $\dot{Q}(\boldsymbol{r})$ scattering remains at the atomic defects along the graphene boundaries, the location of the $\dot{W}(\boldsymbol{r})$ elastic scattering can be shifted towards the bulk of the sample by electrostatics. This finding provides an important insight into two puzzling observations.

The first observation is that the $\dot{W}(\boldsymbol{r})$ contour in Figs. 2a,b is shifted significantly into the bulk in the corner region of the sample. The reason being that the hole accumulation at graphene boundaries has a stronger and longer-range effect at sample corners and in particular in the protrusion area due to the combined electrostatic contribution from multiple boundaries. As a result, the inner nontopological channels are driven deeper into the bulk "cutting the corners", similar to the effect of accumulating PG in Figs. 3n,o.

The second puzzle pertains to the question of the equilibration length. When the QH channels are well equilibrated, no work can be performed since the channels are at the same electrochemical potential [6]. If the equilibration length is short compared to the sample size, the work can be performed only near the contacts (hot spots) where the channels are out of equilibrium, and hence the $R_{xx}(\boldsymbol{r})$ signal will be present only near the contacts and will decay rapidly away from them. The fact that we observe $R_{xx}(\boldsymbol{r})$ and $\dot{W}(\boldsymbol{r})$ contours along the entire length of the sample edges (e.g. Figs. 1e, 2b, Movie M2) shows that the equilibration length of the nontopological channels is larger than the sample size. This conclusion, however, seems to be at odds with the presence of QH plateaus (Fig. S5b), since no conductance quantization is expected in absence of channel equilibration [6,21,37,45–47]. The above observation that the inner nontopological channels "cut" the corners and protrusions can resolve the puzzle by noting that the ohmic contacts (and voltage contacts in particular, Figs. S1-S3) are connected to the Hall bar through narrow graphene protrusions. As a result, the ohmic contacts are apparently equilibrated predominantly with the outermost downstream channels, while the innermost upstream channels do not enter the protruding regions (similar to Fig. 3o). In this situation, very weak equilibration between the outermost and innermost channels can on one hand give rise to relatively well pronounced QH plateaus, while on the other hand a depleting tip can still induce significant $\delta \dot{W}(\boldsymbol{r})$ and $R_{xx}(\boldsymbol{r})$ by locally enhancing the equilibration between the channels. We therefore conclude that the counterpropagating pair of nontopological channels is mostly in the weak tunneling regime, but the tip potential can locally induce a strong coupling between the two channels as shown schematically in Fig. S11b.

### SI7. Additional examples of plunger gate modification of edge reconstruction in samples *B* and *C*

The control of edge accumulation by $V_{pg}$ presented in Fig. 3 and Movie M3 is observed in samples *B* and *C* at various values of $V_{bg}$. Figure S7 shows an additional example in sample *B* at $V_{bg} = -1.0$ V slightly away from the $\nu = -2$ plateau ($\nu = -1.15$, see Fig. S5c) and at higher $V_{tg} = 8$ V, revealing behavior very similar to Movie M3 but with slightly larger features. Figure S8 presents an example from sample *C* in the zeroth LL ($\nu \cong -1$) and low $V_{tg} = 2$ V. In this case only nontopological pairs of edge channels are present along the edges due to charge accumulation, as presented schematically in Figs. S8a, d, and g. The low $V_{tg}$ acts only as a small perturbation to the $\dot{W}(\boldsymbol{r})$ processes resulting in narrow $R_{xx}(\boldsymbol{r})$ features revealing the loci of carrier backscattering. Remarkably, even such small $V_{tg}$ acting on nanoscale regions of the sample, gives rise to an increase in the global $R_{xx}$ by almost 20 k$\Omega$. This implies that in the zeroth LL the bulk is highly resistive and it is shunted by the conductance of the nontopological edge channels. Consequently, cutting off these channels locally by the tip potential gives rise to a very large increase in $R_{xx}$.

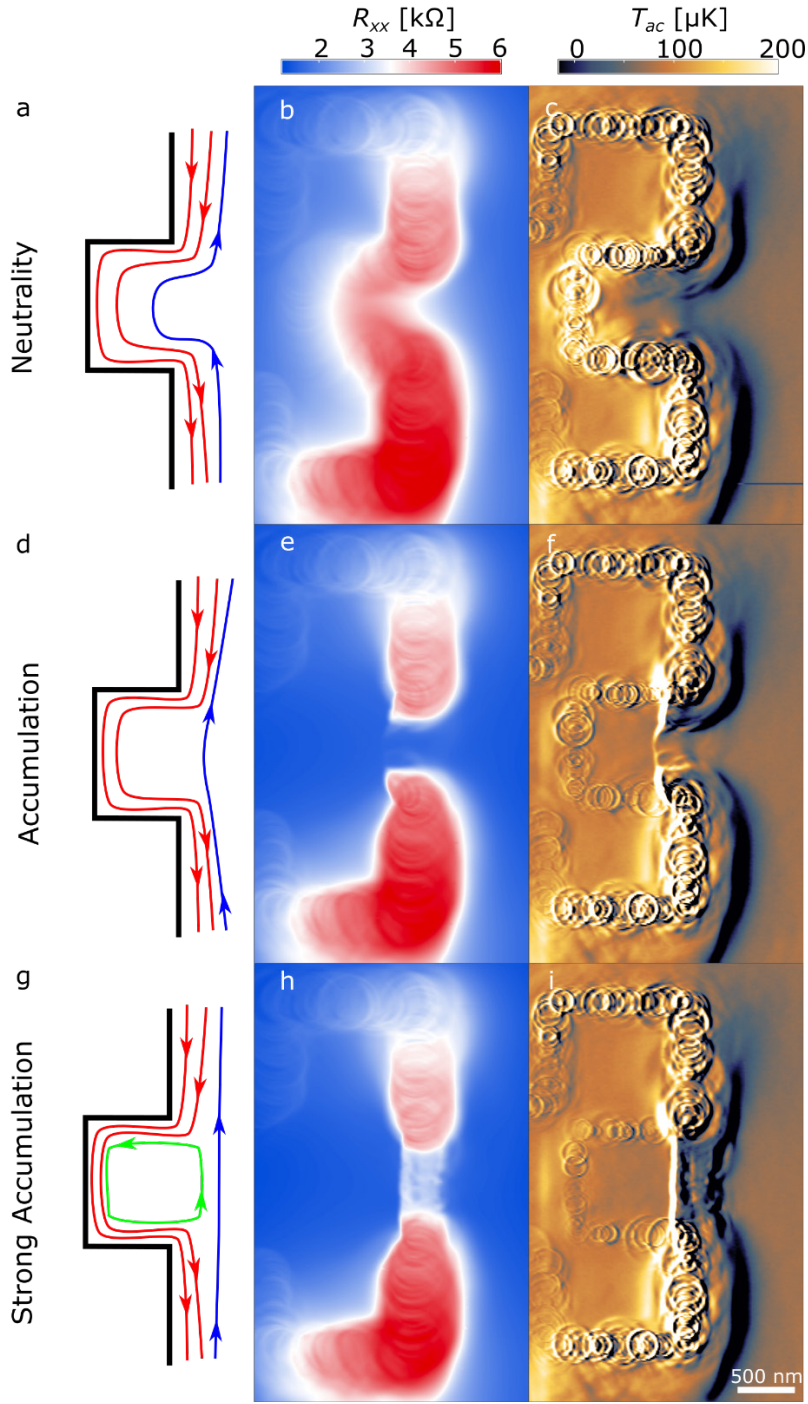

**Figure S7 | Visualization of work and entropy generating processes in sample B. a-c**, Trajectories of QH edge channels (**a**), $R_{xx}(\boldsymbol{r})$ (**b**), and $T_{ac}(\boldsymbol{r})$ (**c**) at $V_{bg} = -1.0$ V ($\nu = -1.15$), $V_{tg} = 8$ V, $V_{pg} = -0.27$ V (neutral PG), and $I_{dc} = 1.75$ μA. **d-f**, Same as (a-c) but $V_{pg} = -0.6$ V (hole accumulating PG). **g-j**, Same as (a-c) but $V_{pg} = -1.4$ V (strongly hole accumulating PG). Images acquired in the dashed red area in Fig. S2.

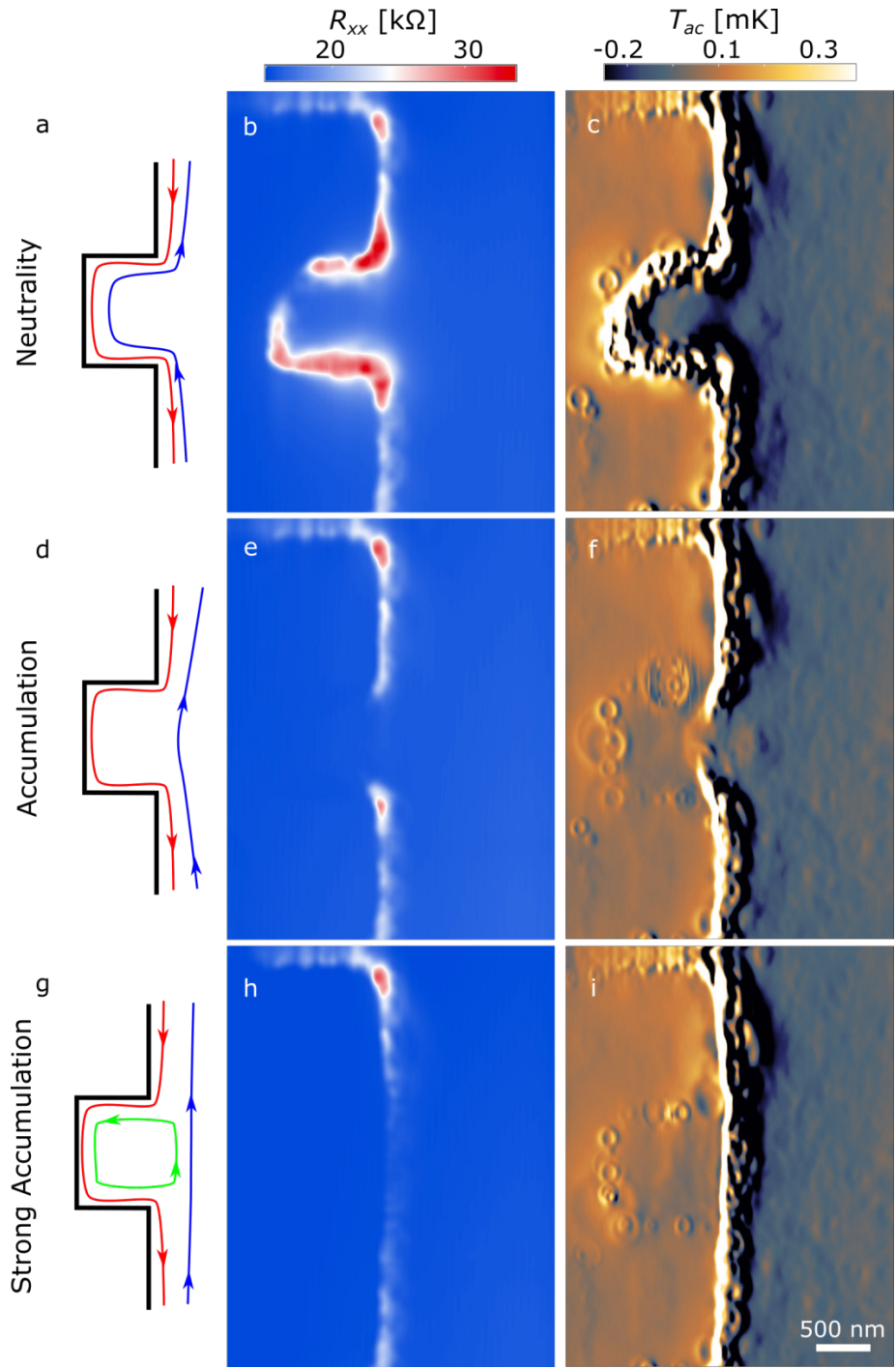

**Figure S8 | Visualization of work and entropy generating processes in sample C. a-c**, Trajectories of QH edge channels (**a**), $R_{xx}(r)$ (**b**), and $T_{ac}(r)$ (**c**) at $V_{bg} = -2.2$ V ($\nu = -0.77$), $V_{tg} = 2$ V, $V_{pg} = -0.3$ V (neutral PG), and $I_{dc} = 0.96$ µA. **d-f**, Same as (a-c) but $V_{pg} = -0.9$ V (hole accumulating PG). **g-j**, Same as (a-c) but $V_{pg} = -2.3$ V (strongly hole accumulating PG). Images acquired in the dashed red area in Fig. S3.

## SI8. Cutting off the nontopological edge channels by plunger gate $V_{pg}$

The enhanced conductivity of the nontopological pairs of channels due to hole accumulation provides low resistance paths for the current flow. This accumulation, however, can be locally depleted by the plunger gate with $V_{pg} > -0.23$ V, as indicated by transport measurements in Fig. S5d. In this case, the nontopological pairs are cut off (Fig. S9a), thus causing a pronounced increase in the global $R_{xx}$. Interestingly, in this situation the current that is carried by the nontopological channels is partially forced to flow through the bulk in the cut off segment. A depleting $V_{tg}$ increases the local bulk resistivity under the tip thus enhancing $R_{xx}(r)$ as observed by the diffused red blob in Fig. S9b and Movie M3 revealing the current path through the bulk. Note that the topologically protected channel remaining in the depleted region (red in Fig. S9a) still carries current, however, the resulting potential drop that develops across the plunger gate region imposes a parallel partial conduction through the bulk. The current that flows in the topological channel, however, cannot be visualized by $R_{xx}(r)$ because the downstream flowing carriers there cannot backscatter to another channel and do not perform work. These carriers, however, can still lose their excess energy by phonon emission at the atomic defects giving rise to the $\dot{Q}$ rings along the graphene boundaries as observed in Fig. S9c.

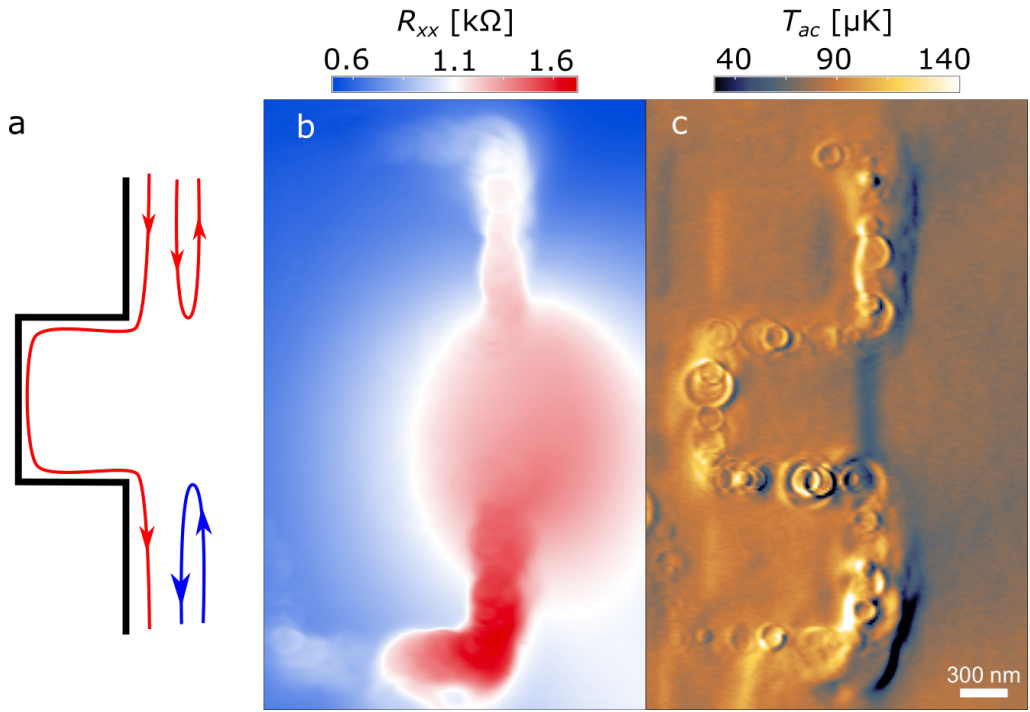

**Figure S9 | Visualization of bulk current flow upon cutting off nontopological channels. a**, Schematic trajectories of QH edge channels with the nontopological pair of channels cut off by the hole-depleting plunger gate. **b**, Scanning gate $R_{xx}(r)$ image in sample B at $V_{bg} = -1.2$ V, $V_{tg} = 3$ V, $V_{pg} = -0.1$ V, and $I_{dc} = 1.75$ µA, revealing current flowing through the bulk the in the cut-off region (diffused red blob). **c**, Simultaneously acquired $T_{ac}(r)$ showing $\dot{Q}$ rings along the graphene boundaries due to nonlocal dissipation. The images were acquired in the dashed red area in Fig. S2.

Since the nontopological channels are cut off, this case provides an insight into work and dissipation that should occur in their absence. Figure S9b shows that the carriers tunnel between the edge states through the bulk as expected in the QH plateau transition regions. When the nontopological edges are present, however, they shunt the bulk by providing low-resistance paths for carrier backscattering and hence hardly any work and dissipation are observed in the bulk of the sample even in the plateau transition regions.

### SI9. Movie M4 – Current dependence

The tip-induced change in the local work generation rate is given by $\delta \dot{W}(r) = (R_{xx}(r) - R_0)I^2$, where $R_0$ is the background resistance in absence of the tip. Remarkably, the tip induced resistance can be extremely large, $R_{xx}(r) \gg R_0$, with $R_{xx}(r) - R_0$ reaching several kΩ and up to 20 kΩ in the zeroth LL (Fig. S8). Despite its very large value we find that $R_{xx}(r)$ is essentially current independent as demonstrated in Movie M4 acquired in Sample B at $V_{bg} = -1.0$ V, $V_{pg} = -2.0$ V, and $V_{tg} = 8.0$ V. Here the ac current $I_{ac}$ is varied by over more than two orders of magnitude from 10 nA to 1.4 µA with only minor change in $R_{xx}(r)$. The current independent $R_{xx}(r)$ implies that the resulting $\dot{W}(r)$ and the nonlocal $\dot{Q}(r)$ increase quadratically with $I_{ac}$. Indeed, the second harmonic thermal signal $T_{2f}$ in Movie M4 is below our sensitivity at low currents and grows quadratically with the current (see SI2 for explanation of the $T_{2f}$ signal). Note that the sharp $\dot{Q}$ rings in $T_{2f}(r)$ images at elevated currents are district from the $\dot{W}(r)$ arc-like patterns visible both in $T_{2f}(r)$ and $R_{xx}(r)$ scans.

This finding of linear $V_{xx}(r)$ over more than two orders of magnitude in current (constant $R_{xx}(r)$) shows that the observed dissipation mechanism and the suppression of the apparent topological protection is a linear effect. This is in stark contrast to the common current-induced breakdown of the QH state which is a highly nonlinear phenomenon and is generally attributed to the onset of dissipation in the bulk [4,48–50]. The nontopological edge channels effectively act as "linear" resistors along the edges of the sample shunting the interior both in the compressible and incompressible bulk states.

### SI10. Hole accumulation at the graphene edges for *n*-doped bulk

Edge charge accumulation causes formation of nontopological pairs of channels that provide a low resistance path for current flow. These nontopological pairs can be cut off by a depleting plunger gate leading to increase in $R_{xx}$. In the case of hole accumulation this is demonstrated by applying a positive $V_{pg}$, while a negative $V_{pg}$ does not affect $R_{xx}$ substantially because increasing local accumulation only lowers the local resistance, which is already relatively low (Fig. S5d). Figure S5c indeed shows that for negative $V_{bg}$ (*p* doped bulk) a depleting (positive) $V_{pg}$ increases $R_{xx}$. If the edge accumulation would be solely caused by backgate electrostatics [7,20,51,52], the situation would be inverted for *n* doped bulk, namely a negative $V_{pg}$ would deplete the electron accumulation along the edges thus increasing $R_{xx}$.

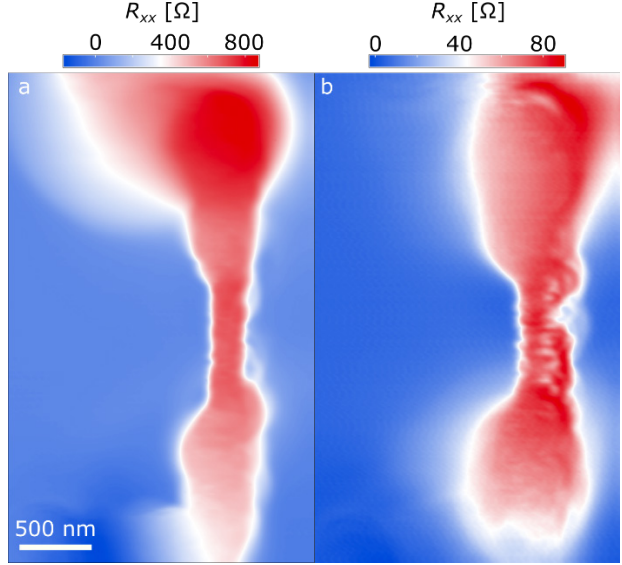

**Figure S10 | Demonstration of hole edge accumulation in *n*-doped bulk. a**, Scanning gate $R_{xx}(r)$ imaging of sample B (in the red dashed area in Fig. S2) in the vicinity of *n*-doped $\nu = 2$ plateau ($V_{bg} = 0.12$ V, $\nu = 2.07$, $V_{pg} = -2$ V, see Fig. S5c for transport) using a positive $V_{tg} = 6$ V. The depletion of the hole accumulated edges by the positive $V_{tg}$ cases increase in $R_{xx}(r)$ similar to the *p*-doped bulk case. **b**, Same as (a) in the vicinity of $\nu = 6$ plateau ($V_{bg} = 1.475$ V, $\nu = 5.98$). In both images $I_{dc} = 1.75$ µA.

Figure S5c shows, however, that it is not the case and $R_{xx}$ is increased by a positive $V_{pg}$ even in the *n*-doped region (for not too high $V_{bg} \lesssim 2.3$ V). This implies that for moderate *n* doping of the bulk, the edges still remain *p* doped, as confirmed microscopically in Fig. S10. Here the $R_{xx}(r)$ scans for *n*-doped bulk show that similarly to the case of *p*-doped bulk, a positive, rather than negative, $V_{tg}$ increases the resistance along the edges. This hole edge accumulation [53] is clearly resolved in the vicinity of $\nu = 2$ and 6 plateaus as shown in Figs. S10a,b.

### SI11. Demonstration of the elastic $\dot{W}$ scattering and nonlocal heating

We present here a more detailed evidence that $\dot{W}(r)$ process is predominantly elastic. For this we first summarize the effect of the tip potential $V_{tg}$ which is crucial for revealing the described phenomena. At flat band conditions $V_{tg} = V_{tg}^{FB} \cong 0$ V the tip has no influence on the sample. In this case $R_{xx}(r)$ is fixed independent of the tip position (Fig. S11a) and therefore the $\dot{W}(r)$ processes cannot be imaged (Fig. S12a). Nonetheless, the $T_{dc}(r)$ signal due to $\dot{Q}(r)$ processes is present, but the phonons, even though being emitted predominantly at resonant states at atomic defects, propagate ballistically throughout the sample. As a result, the $T_{dc}(r)$ profiles are smooth (Fig. S12b) and hence the atomic-scale $\dot{Q}(r)$ sources cannot be resolved individually.

Upon applying a finite $V_{tg}$, however, both the $\dot{Q}(r)$ and $\dot{W}(r)$ processes can be clearly identified. The individual $\dot{Q}(r)$ sources are revealed through the formation of the temperature rings around them (e.g. Fig. S7), which reflect the loci of the tip positions at which the tSOT potential $V_{tg}$ brings the localized resonant electronic states of the defects to the Fermi energy [11]. Similarly, applying a small positive $V_{tg}$ allows imaging the locations at which $\dot{W}(r)$ is present and hence revealing the locations of the nontopological channels by shifting them slightly closer to each other (Fig. S11b), thus enhancing the local elastic tunneling rates between them by $\delta\dot{W}(r)$ and increasing the $R_{xx}(r)$. The observed rich patterns of $\delta \dot{W}(\boldsymbol{r})$ reflect the intricate trajectories of the QH channels and the variations in the local separation between them due to electrostatic disorder. In particular, it reveals the nontopological channels at the inner edge of the plunger gate in Fig. S12c. A further increase of the depleting $V_{tg}$ can entirely cut off the nontopological pairs (Fig. 11c) and even induce an *n*-doped region under the tip (Fig. 11d).

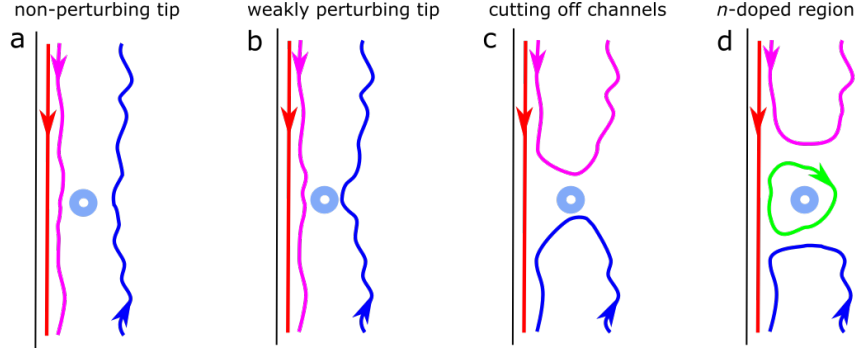

**Figure S11 | Effect of the tip potential $V_{tg}$ on the QH channels with *p*-doped edge accumulation**. **a-d**, Schematic trajectories of the edge channels upon increasing $V_{tg}$. **a**, Non-perturbing tip at flat band conditions $V_{tg} = V_{tg}^{FB} \cong 0$ V. **b**, Weakly perturbing $V_{tg}$ slightly reducing the edge hole accumulation and shifting the nontopological QH channels closer to each other. **c**, Stronger depleting $V_{tg}$ cutting off the nontopological pair of channels. **d**, Higher $V_{tg}$ forming an *n*-doped region under the tip.

Throughout the paper we refer to the $\dot{W}$ process of carrier tunneling between the channels as a purely elastic process with no local phonon emission, in which case all the $\dot{Q}$ processes are nonlocal (Fig. 2d). One can also consider a higher-order inelastic tunneling between the channels, in which a phonon is emitted concurrently with tunneling, resulting in a local $\dot{Q}$ at the tunneling location. We can discern the two cases by considering the perturbation induced by the tip potential. A weakly perturbing tip has two effects: enhancing backscattering $\delta \dot{W}(\boldsymbol{r})$, thus revealing the locations of $\dot{W}(\boldsymbol{r})$ processes through $R_{xx}(\boldsymbol{r})$ imaging (Fig. S12c), and enhancing heating (either local or nonlocal) as a result of the enhanced $\delta \dot{W}(\boldsymbol{r})$. Figures 2, S7 and Movie M3 clearly demonstrate that the $\dot{Q}$ rings at atomic defects along the graphene boundaries reflect nonlocal heating. However, the observed enhanced temperature signal along the QH channels (e.g. on the bulk-side edge of the PG in Fig. S12d) could reflect either local heating due to higher-order inelastic carrier tunneling between the QH channels or a nonlocal heating due to phonons emitted at remote locations causing overall temperature increase detected as an enhanced $T_{dc}(\boldsymbol{r})$ at the instantaneous position of the tip $\boldsymbol{r}$. These two possibilities are hard to distinguish by inspecting only the tip-perturbing images like Fig. S12d. A non-perturbing tip, in contrast, performs only one function – imaging the unperturbed temperature distribution. If the tunneling between QH channels is elastic then the heating is nonlocal and thus the maximum of $T_{dc}(\boldsymbol{r})$ should occur along the graphene boundaries where the phonons are emitted at the atomic defects regardless of where $\dot{W}(\boldsymbol{r})$ occurs. If, on the other hand, the tunneling is inelastic, a peak in $T_{dc}(\boldsymbol{r})$ should occur along the $\dot{W}(\boldsymbol{r})$ contours. Usually the $\dot{W}(\boldsymbol{r})$ contours are located close to the graphene boundaries, but near the sample corners they can be significantly shifted towards the bulk (Fig. 2a,b) or, alternatively, we can shift them in a controllable manner by the PG (Fig. S12).

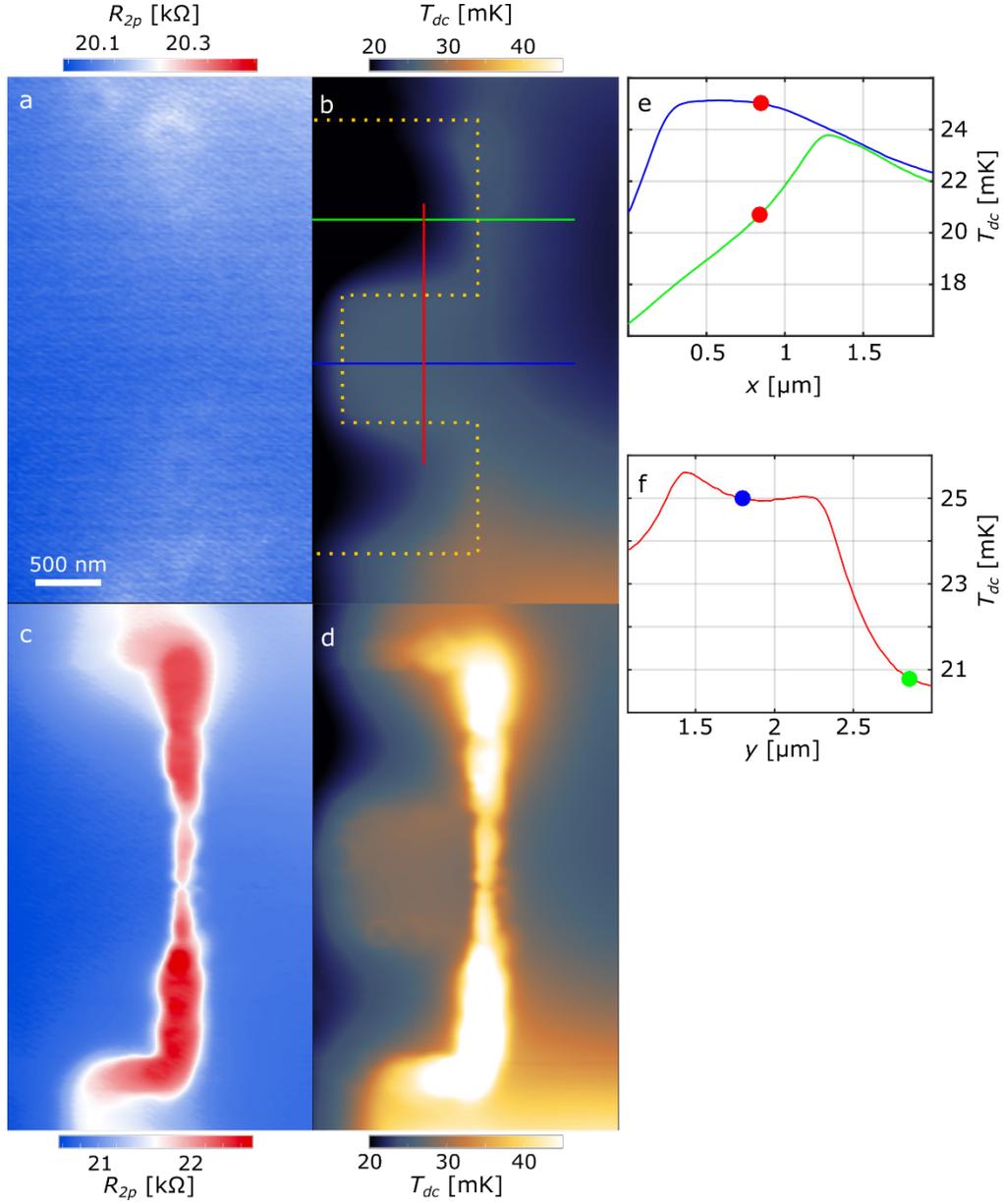

**Figure S12 | Demonstration of elastic tunneling by comparing perturbing and non-perturbing tip potential in sample B. a**, Two probe $R_{2p}(r)$ in the case of non-perturbing $V_{tg} = 0.05$ V $\cong V_{tg}^{FB}$ showing essentially constant $R_{2p}(r)$. **b**, The corresponding $T_{dc}(r)$ shows the current induced temperature variation in the sample unperturbed by the tip at $V_{bg} = -1.1$ V ($\nu = -1.44$), $V_{pg} = -2$ V, and $I_{dc} = 1.75$ μA. The enhanced temperature at the bottom-right corner is caused by heat diffusion from the hot spot at the nearby current contact. **c**, $R_{2p}(r)$ for $V_{tg} = 3$ V revealing the location of $\dot{W}(r)$ processes by perturbing the local work by $\delta\dot{W}(r)$ through enhanced backscattering. **d**, The corresponding $T_{dc}(r)$ showing the temperature map mimicking the $R_{2p}(r)$ signal caused by the enhanced nonlocal heat release $\dot{Q}$ due to tip-induced $\delta\dot{W}(r)$. **e**, Horizontal line cuts of $T_{dc}(r)$ along the green and blue lines in (b). The green data show a peak at the graphene boundary (dashed yellow line in (b)) followed by a slowly decaying tail into the bulk, while the blue data display no peak at the inner edge of the PG showing that the $\dot{W}(r)$ process there is elastic. **f**, Vertical line cut through the protrusion region showing peaks at the graphene boundaries with overlapping tails in the middle. The color dots are the intersection points of the lines.

Figures S12e,f present three $T_{dc}(r)$ profiles along the color lines in Fig. S12b for the case of a non-perturbing tip. The green profile shows that $T_{dc}(r)$ is maximal along the graphene boundaries with a slowly decaying tail into the bulk of the sample due to ballistic phonon propagation. The red and blue profiles show that the slow tails of $T_{dc}(r)$ originating from the three boundaries overlap, resulting in a plateau-like profile in the sample protrusion region. The key observation, however, is that the blue profile in Fig. S12e shows no peak in $T_{dc}(r)$ at the location of the $\dot{W}(r)$ contour on the bulk side of the PG as revealed in Figs. S12c,d. These results demonstrate that the $\dot{W}(r)$ scattering is predominantly elastic and that the $\dot{Q}(r)$ dissipation is predominantly nonlocal occurring at the atomic defects at graphene boundaries.

### SI12. Movie M5 – Effects of the tip potential $V_{tg}$ on $R_{xx}(r)$

Movie M5 shows an example of the evolution of $R_{xx}(r)$ upon varying $V_{tg}$ in sample *B* at $V_{bg} = -1$ V ($\nu = -1.15$), $V_{pg} = -0.22$ V (neutral plunger gate), and very low current of $I_{ac} = 10$ nA. A negative $V_{tg}$ causes accumulation of holes under the tip, but this has no observable effect. This is because hole accumulation is already present along the edges and increasing this accumulation in a very small region does not influence (decrease) the backscattering appreciably (in sharp contrast to depleting the hole accumulation) consistent with transport data in Fig. S5d. As $V_{tg}$ is increased to small positive values, the induced depletion of the hole accumulation causes compression of the counterpropagating channels (Fig. 11b), resulting in enhanced backscattering and appearance of corresponding features in $R_{xx}(r)$ which reveal the locations of the most dominant scattering sites. When $V_{tg}$ becomes sufficiently large (e.g. 1.75 V) to cut off the counterpropagating pairs of channels (Fig. 11c), the enhanced $R_{xx}(r)$ becomes visible along the entire edge of the sample where the nontopological channels are present, displaying a highly disordered structure. For $V_{tg} \gtrsim 3$ V arc-like features are formed which increase in diameter and become very fine upon further increase of $V_{tg}$. In this case an *n*-doped pocket is formed under the tip (Fig. 11d). At high $V_{tg}$ this pocket will contain a number of LLs with edge channels strongly compressed against the steep edge potential, apparently causing enhanced backscattering between the channels by the resonant states at the individual atomic defects. The arcs are very fine at the applied low current of 10 nA and become more blurred at higher currents (compare to Fig. S7 at 1.75 μA).